\documentclass[aps,reprint,superscriptaddress,amsmath,amssymb,graphicx]{revtex4-2}
\usepackage{graphicx}
\usepackage[caption=false]{subfig}
\usepackage{rotating}
\usepackage{adjustbox}
\usepackage[english]{babel}
\usepackage{booktabs}
\usepackage[utf8]{inputenc}
\usepackage[T1]{fontenc}
\usepackage{amsmath} 
\usepackage{placeins}
\usepackage{hyperref}
\hypersetup{
  colorlinks=true,
  linkcolor=black,
  citecolor=black,
  urlcolor=black
}

\begin{document}
\raggedbottom

\title{Guiding Peptide Kinetics via Collective-Variable Tuning of Free-Energy Barriers}



\author{Alexander Zhilkin}
\affiliation{The Wolfson Department of Chemical Engineering,
Technion - Israel Institute of Technology, Haifa 32000, Israel}

\author{Muralika Medaparambath}
\affiliation{The Wolfson Department of Chemical Engineering,
Technion - Israel Institute of Technology, Haifa 32000, Israel}
\affiliation{Faculty of Mathematics,
Technion - Israel Institute of Technology, Haifa 32000, Israel}

\author{Dan Mendels}
\email{danmendels@technion.ac.il}
\affiliation{The Wolfson Department of Chemical Engineering,
Technion - Israel Institute of Technology, Haifa 32000, Israel}




\begin{abstract}
While recent advances in AI have transformed protein structure prediction, protein function is also strongly influenced by the thermodynamic and kinetic features encoded in its underlying free-energy surface. Here, we propose a data-efficient framework for engineering protein conformational kinetics by rationally reshaping free-energy landscapes to control transition rates. Built on the Collective Variables for Free Energy Surface Tailoring (CV-FEST) framework, the approach is validated on point mutations of the miniprotein Chignolin. The framework relies on Harmonic Linear Discriminant Analysis (HLDA)-based collective variables (CVs) constructed from short molecular dynamics trajectories confined to metastable folded and unfolded basins, requiring only limited local sampling rather than exhaustive rare-event simulations. Notably, the HLDA CV derived solely from the wild-type system provides residue-level scores that predict whether mutations at specific positions are likely to accelerate or slow unfolding transitions. Furthermore, the leading HLDA eigenvalue associated with the derived CV, a quantitative measure of the one-dimensional statistical separation between folded and unfolded ensembles, is significantly correlated with transition rates across mutations. Together, these results suggest that mutation-dependent kinetic effects can be inferred from minimal in-basin sampling, providing a practical route for guiding peptide and protein engineering through collective-variable design, free-energy surface engineering, and data-efficient molecular simulation.

\end{abstract}


\maketitle 

\section{Introduction}

Protein conformational dynamics are central to biological function and directly influence processes such as molecular recognition, signal transduction, and drug delivery \cite{dynamic-personalities-of-proteins, boehr-2009}. By modulating how proteins engage binding partners and form protein–protein interactions, these dynamics tune binding mechanisms and interaction pathways and play an increasingly important role in peptide and protein-based therapeutics \cite{keskin2016}. Importantly, beyond their underlying thermodynamics, functional outcomes often also depend on the rates at which proteins interconvert between conformational states.

Conformational dynamics can influence ligand residence times, shape allosteric signaling, and regulate catalytic efficiency. In the context of drug delivery, these motions directly affect the dissociation rate ($k_{\text{off}}$), where slower unbinding prolongs target engagement and is frequently associated with improved therapeutic efficacy \cite{ligand:binding}. As a result, understanding and controlling conformational transition kinetics is a key challenge in peptide and protein design.

Predicting kinetic observables associated with conformational transitions remains a major challenge. Computational approaches to mutation effects on peptides and proteins are often framed around two related but distinct questions: thermodynamic stability and kinetic transitions; here, we focus exclusively on the latter \cite{rate-prediction-review}. Most existing predictors rely on static sequence- or structure-derived inputs. Early sequence-based models, such as FOLD-RATE \cite{lin2006foldrate}, SWFoldRate \cite{wang2013swfoldrate}, FoldRate \cite{2009foldrate}, SeqRate \cite{zou2010seqrate}, PRORATE \cite{prorate}, and Pred-PFR \cite{wei2014predpfr}, estimate folding rates from fixed sequence descriptors such as composition and window-based features. More recent supervised predictors incorporate structural information and curated experimental annotations; for example, K-Fold \cite{kfold} and FRTpred \cite{kang2022frtpred} infer folding rates and, in some cases, folding type or kinetic order from experimentally measured datasets.

Dedicated predictors of mutation-induced rate changes have also been explored. However, mutation-specific kinetics prediction remains limited by the availability of experimentally measured mutant folding and unfolding rates and by dataset imbalance across mutation types, which can restrict model training and generalization. Although kinetic databases such as K-Pro \cite{kpro} and KineticDB~\cite{kineticdb} exist, the coverage of experimentally measured folding and unfolding rates, particularly for mutant variants, remains modest relative to the needs of data-intensive modeling. Consequently, current approaches remain largely dependent on static inputs or curated training datasets, limiting generalization to proteins or mutation types that are underrepresented in existing measurements and reflecting a broader challenge for machine-learning approaches in data-scarce scientific settings \cite{recent-protein-mech,salman-data-scar}.

Molecular dynamics simulations can, in principle, provide direct access to conformational kinetics. However, exhaustive sampling of rare transitions is often computationally prohibitive; even for small miniproteins such as Chignolin, first-passage times can extend well beyond the microsecond timescale~\cite{trajectory}. Enhanced sampling approaches~\cite{metadynamics:to:dynamics, McCarty-parrinello-2017, Invernizzi-parinello-2020} address this limitation by accelerating rare events while preserving access to unbiased kinetics through carefully controlled bias deposition. However, these methods can still require substantial computational resources and manual intervention, particularly in high-throughput settings. 

To circumvent these limitations the Collective Variables for Free Energy Surface Tailoring (CV-FEST) framework \cite{cv-fest:dan} proposes identifying low-dimensional, physically interpretable CVs that capture a system’s slow modes and govern rare barrier-crossing events. The central concept is to modulate kinetics by deliberately reshaping free-energy barriers along these CVs, rather than relying on large training datasets or extensive rare-event sampling.  

Within this framework, CV construction methods such as Harmonic Linear Discriminant Analysis (HLDA) provide a practical and physically grounded approach for constructing low-dimensional CVs. This approach requires only limited training data obtained from short simulations confined to the metastable states of interest, without the need to directly sample transition events. It constructs CVs as linear combinations of user-defined descriptors, yielding physically interpretable results in which descriptors with larger weights correspond to greater contributions to the system’s slow dynamics.

Here, we build on CV-FEST to examine and predict how point mutations alter conformational kinetics through changes in barrier heights. We demonstrate the methodology on the extensively studied yet kinetically nontrivial Chignolin peptide, a canonical benchmark for folding and rare-event kinetics. Beyond their role as convenient model systems, peptides are also of broad biological and practical interest: short peptides mediate a substantial fraction of protein--protein interactions (15\%--40\%) \cite{furman2013} and play central roles in molecular recognition, signaling, and regulation \cite{CUNNINGHAM201759}. Consequently, understanding how point mutations reshape peptide free-energy landscapes and conformational kinetics is relevant not only for methodological development, but also for a wide range of biological and biomedical applications.

We find that a CV constructed using HLDA from wild-type (WT) simulations alone provides residue-level guidance for mutation design. The dominant eigenvector assigns interpretable weights to the underlying descriptors, thereby identifying residues whose perturbation is more likely to accelerate or slow unfolding kinetics. The corresponding eigenvalues computed for specific amino acid substitutions provide a quantitative measure of the separation between folded and unfolded ensembles in the mutants, which we find to show significant correlation with the mean MFPTs across individual point mutations. Together, these results indicate that WT-derived residue importance and mutation-specific state separation can capture consistent underlying kinetic trends, paving the way for a data-efficient strategy for screening mutations that modulate unfolding rates. 
\section{Methods}
\subsection*{CV-FEST framework for free-energy surface engineering}
This study is conducted within the Collective Variables for Free Energy Surface Tailoring (CV-FEST) framework, which provides a systematic route to modifying functionality in systems governed by rare conformational events~\cite{cv-fest:dan}. The central assumption of this framework is that the key thermodynamic and kinetic features of such systems are encoded in a low-dimensional representation of the FES, expressed in terms of CVs that capture the dominant slow degrees of freedom. By projecting the dynamics onto this reduced space, the relevant information governing state stability and transition barriers is condensed into a small set of parameters, allowing free-energy differences and barrier heights to be manipulated in a controlled manner without the need to explicitly sample full transition pathways.

CVs are defined as functions of the system’s microscopic coordinates, $\mathbf{s}(\mathbf{R})$. The probability distribution along the CV is given by
\begin{equation}
P(\mathbf{s}) = \int d\mathbf{R}\,\delta\!\left[\mathbf{s}(\mathbf{R}) - \mathbf{s}\right] P(\mathbf{R}),
\end{equation}
where $P(\mathbf{R})$ denotes the Boltzmann probability distribution and $\delta$ is the Dirac delta function. The corresponding free-energy surface (FES) with respect to the chosen CV follows as
\begin{equation}
F(\mathbf{s}) = -k_B T \log P(\mathbf{s}),
\end{equation}
where $k_B$ is Boltzmann’s constant and $T$ is the system temperature.

Perturbations to the system modify the underlying probability distribution $P(\mathbf{s})$ and thereby reshape the FES, in particular the free-energy difference between metastable states and the barrier heights associated with the rare conformational transition of interest. 
Previous applications of CV-FEST focused on continuous tuning of system interactions or forces along the identified CVs, providing a controlled setting for probing structure-function relationships~\cite{cv-fest:dan, hlda:dan}. Here, we extend this framework to a more realistic and experimentally relevant scenario in which perturbations arise from discrete point mutations.

\subsection*{Collective variable construction via Harmonic Linear Discriminant Analysis}

Within the CV-FEST framework, we opt to use Harmonic Linear Discriminant Analysis (HLDA) as the central tool for constructing data-efficient and interpretable CVs that capture the relevant slow modes of the peptide, following the formulation introduced by Mendels \emph{et al.}~\cite{mendels:local-fluctuations, hlda:dan, folding-small-protein, dan:discriminant, dan:crystal, dan:blind}. This approach serves as a convenient practical realization of the CV-FEST philosophy by expressing the CV as a linear combination of physically motivated descriptors, trained solely on short simulations confined to metastable basins.

The descriptor space is defined by backbone distance descriptors between residue pairs of the peptide, as illustrated in Fig.~\ref{fig:backbone}(a). To improve numerical stability and avoid ill-conditioned covariance matrices, redundant descriptors are removed prior to training. Specifically, we compute the Spearman correlation matrix over all candidate descriptors and iteratively discard one descriptor from any pair with \(\rho > 0.93\), until no such pairs remain. This cutoff is an important calibrated parameter, chosen to balance numerical stability with preservation of descriptor diversity.

HLDA requires estimates of the expectation value vectors $\mu_I$ and covariance matrices $\Sigma_I$ of the descriptor set for each metastable state $I$, here corresponding to the folded ($F$) and unfolded ($U$) ensembles, which are computed from state-restricted, unbiased simulations. The goal is to identify a one-dimensional projection of the descriptor space that maximally separates the folded and unfolded ensembles while minimizing fluctuations within each state. To this end, HLDA determines a projection direction $\mathbf{W}$ by maximizing the ratio between the between-class and within-class scatter matrices. The between-class scatter is
\begin{equation}
\mathbf{S}_b = \sum_I (\mu_I - \bar{\mu})(\mu_I - \bar{\mu})^T
\end{equation}
with $\bar{\mu}$ the global mean, while the within-class scatter is defined via the harmonic average of the state covariances,
\begin{equation}
\mathbf{S}_w = \left( \Sigma_F^{-1} + \Sigma_U^{-1} \right)^{-1}
\end{equation}

For two states, the global mean is
\begin{equation}
\bar{\mu} = \frac{1}{2}\left(\mu_F + \mu_U\right)
\end{equation}
yielding
\begin{equation}
\mathbf{S}_b = \frac{1}{2}\left(\mu_F - \mu_U\right)\left(\mu_F - \mu_U\right)^T
\end{equation}

The eigenvector $\mathbf{W}=\{W_{pq}\}$ associated with the largest eigenvalue $\lambda$ of this construction defines the weights of the resulting CV,

\begin{equation}
s(\mathbf{R}) = \sum_{p<q} W_{pq}\, d_{pq}(\mathbf{R}),
\end{equation}
where $d_{pq}(\mathbf{R})$ denotes the backbone distance between residues $p$ and $q$. The corresponding eigenvalue $\lambda$ provides a scalar measure of the degree of separation between the folded and unfolded ensembles along this direction. 

Conceptually, HLDA formulates CV construction as a classification problem between predefined metastable states, assigning larger weights to descriptors that contribute most strongly to their statistical separation along the constructed CV. Because this projection captures the dominant folded–unfolded slow mode within the chosen descriptor space, descriptors with larger weights are those most strongly associated with this slow conformational transition. Chemical perturbations that modify these descriptors, such as point mutations, are therefore expected to alter the statistical separation between the metastable ensembles along this mode. We therefore hypothesize that mutation-induced changes in separability, quantified by the leading HLDA eigenvalue $\lambda$, may serve as a surrogate measure of changes in the underlying free-energy barrier governing the rare event.

\subsection*{System and state-restricted sampling}
\label{sec:simulation-details}

All calculations are performed on the Chignolin peptide, a ten-residue $\beta$-hairpin that serves as a canonical benchmark for folding and rare-event kinetics. Two metastable conformational states are considered throughout this study: a folded hairpin state and an unfolded ensemble.

To label configurations as folded or unfolded, we use the backbone RMSD to the minimum-enthalpy structure
of the native folded hairpin as a practical structural classifier. We define a folded cutoff $t_F$ and an
unfolded cutoff $t_U$, such that configurations with RMSD $\le t_F$ are assigned to the folded ensemble and
configurations with RMSD $\ge t_U$ are assigned to the unfolded ensemble; configurations with $t_F < \mathrm{RMSD} < t_U$ are excluded to avoid cross-contamination. These thresholds are selected based on qualitative inspection of trajectories, and are used only to define state-restricted ensembles (for HLDA training), not as a reaction coordinate or as the CV employed for kinetic inference. See a more detailed analysis of these values and their impact in the Computational Details section.

To generate the training data used for CV construction via HLDA, we perform short unbiased MD simulations initiated from minimum enthalpy configurations representative of the two basins, with trajectories consisting roughly of 100 ns per state. These simulations are restricted to the folded or unfolded region by construction, and only state-resolved equilibrium fluctuations are used as input for the subsequent CV construction and analysis. Importantly, the CV is learned without requiring any transition frames between the states, consistent with the low-data philosophy of CV-FEST.

\subsection*{Residue-level importance and point mutation selection}

\begin{figure}[t]
\centering
\subfloat[]{%
  \includegraphics[width=0.43\linewidth]{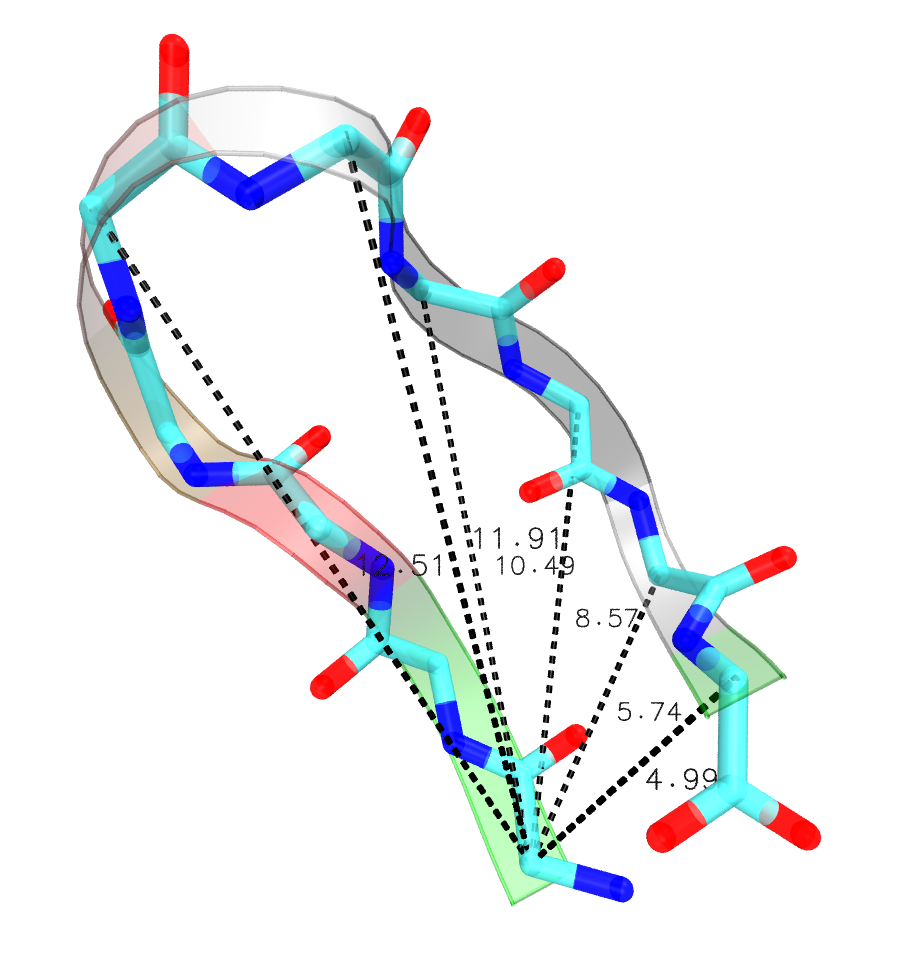}%
  \label{fig:backbone:descriptors}%
}
\hfill
\subfloat[]{%
  \includegraphics[width=0.56\linewidth]{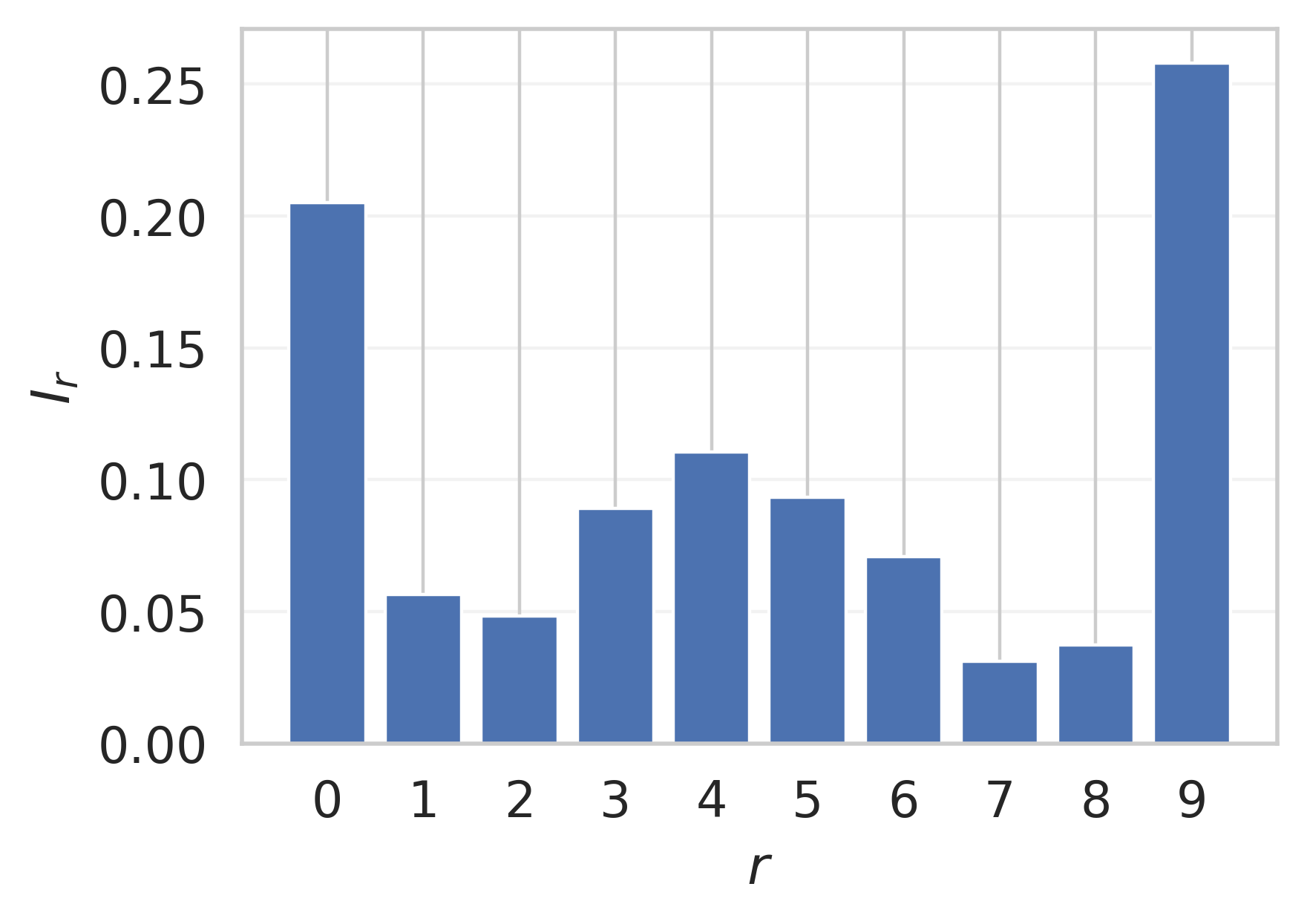}%
  \label{fig:backbone:weights}%
}
\caption{
(a) Illustration of representative backbone-distance descriptors used as input features for HLDA construction.
(b) Aggregated per-residue HLDA weights derived from the leading eigenvector of the WT projection using Eq.~\ref{eq:residue-importance}, shown as normalized values.
}
\label{fig:backbone}
\end{figure}

The projection vector $\mathbf{W}$ assigns a weight to each inter-residue distance $d_{pq}$. Because these weights are defined for residue pairs rather than for individual residues, a residue-level importance measure is obtained by aggregating pairwise contributions by residue. Specifically, the importance of residue $r$ is defined as the average magnitude of all pairwise weights involving that residue.
\begin{equation}
I_r = \frac{1}{|\mathcal{N}_r|} \sum_{q \in \mathcal{N}_r} |W_{rq}|,
\label{eq:residue-importance}
\end{equation}
where $\mathcal{N}_r$ denotes the set of residues paired with residue $r$ in the descriptor set. The resulting normalized scores define a per-residue score profile for the WT peptide, shown in Fig.~\ref{fig:backbone}(b), and are interpreted as a magnitude-based measure of how strongly mutations at a given residue are expected to influence unfolding kinetics.

Residues with large importance scores are predominantly located near the turn and terminal regions of the peptide, consistent with previous studies indicating that Chignolin folds via a turn-directed, edge-to-center “zipping” mechanism in which these regions play a central kinetic role~\cite{rmsd2012,granger-causaility}. This observation motivates the selection of seven residues spanning a range of predicted importance scores. For each selected residue, multiple substitutions are introduced by choosing replacement amino acids with distinct physicochemical properties. This procedure results in four to six mutations per residue and a total of 36 mutants considered in this study.

\subsection*{Kinetic inference via short-time infrequent metadynamics}

To compute conformational transition rates, we employ short-time infrequent metadynamics (ST-iMetaD)~\cite{blumer:2024}, an extension of infrequent metadynamics for estimating the rates of rare events from accelerated simulations. Infrequent metadynamics infers transition rates by rescaling first-passage times using a bias-dependent acceleration factor, under the assumption that escape events from long-lived metastable states follow Poisson statistics. ST-iMetaD improves the efficiency of this inference by basing the rate estimation on short-time transition events, enabling reliable kinetic estimates while allowing more frequent bias deposition.

In these simulations, the previously constructed HLDA CV is used as the biasing coordinate to accelerate unfolding events and enable kinetic estimation. Mean first-passage times are extracted from the resulting trajectories, with full details of the biasing parameters, validation of the exponential survival assumption, and MFPT extraction provided in the Supplementary Information.

\section{Results}

We first analyze the kinetic information encoded in the WT HLDA eigenvector and then show that the corresponding mutation-specific eigenvalue, which quantifies state separation, shows a clear correlation with unfolding kinetics.

\subsection*{Wild--type HLDA weights predict mutational effects on kinetics}

\begin{figure*}[t]
\centering
\subfloat[]{%
  \includegraphics[width=0.49\linewidth]{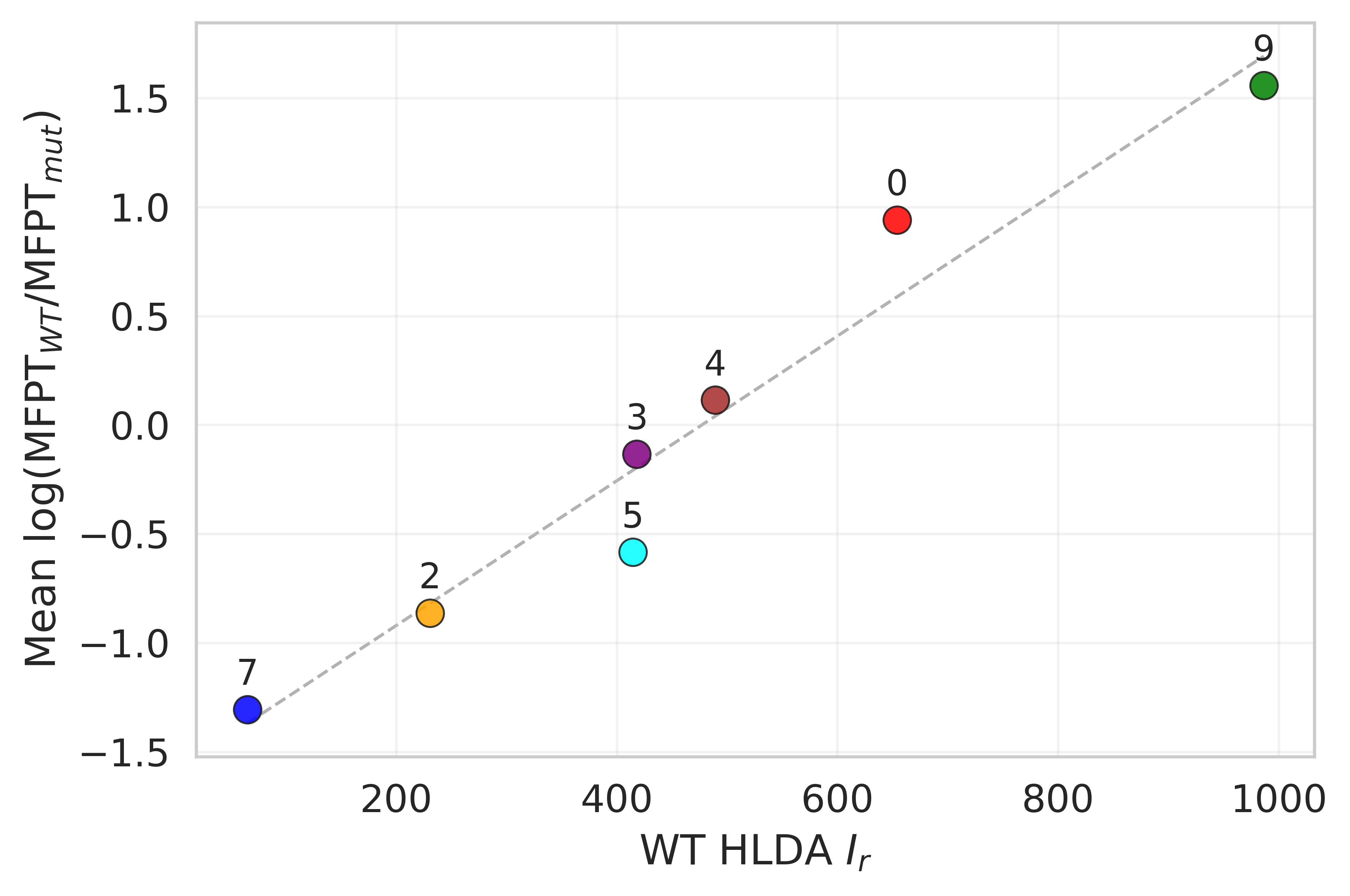}%
  \label{fig:wt-weight-mfpt-scatter}%
}
\hfill
\subfloat[]{%
  \includegraphics[width=0.49\linewidth]{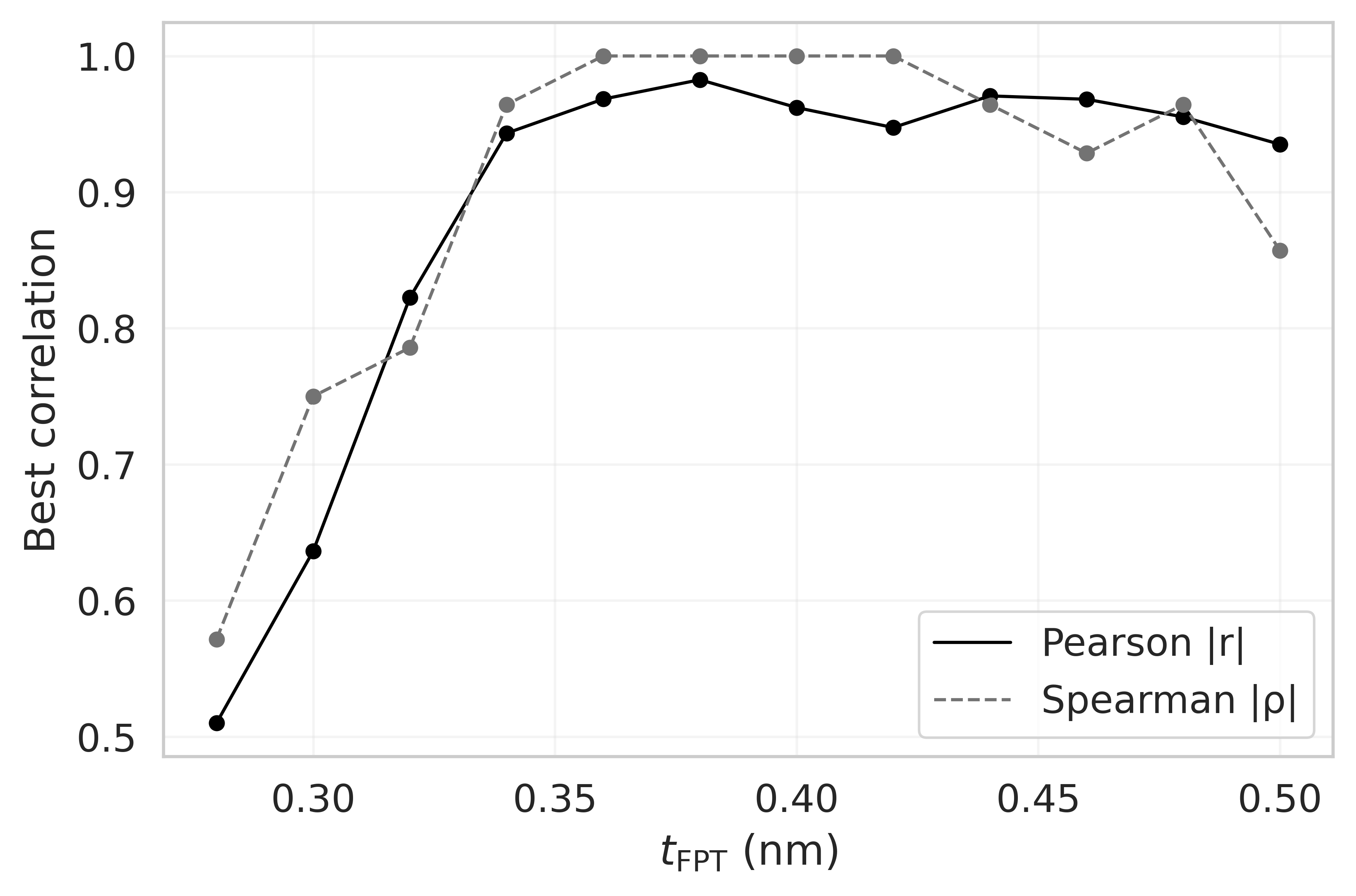}%
  \label{fig:wt-weight-threshold-corr}%
}
\caption{Correlation between WT residue importance and unfolding kinetics.
(a) Aggregated WT residue importance $I_r$ versus mean change in MFPT ($t_{\mathrm{FPT}} = 0.38$) due to mutations at residue $r$, where the mean is taken over a subset of single-point mutations at that position. Pearson $r = 0.96$, $p = 5.4\times10^{-4}$; Spearman $\rho = 1.00$.
(b) Correlation between mean MFPT and aggregated WT residue importance $I_r$ as a function of the first-passage threshold $t_{\mathrm{FPT}}$, demonstrating robustness of the observed relationship across threshold choices.}
\label{fig:wt-weight-mfpt-correlation}
\end{figure*}

We find that the residue-level importance scores $I_r$ (Eq.~\eqref{eq:residue-importance}), derived from the WT HLDA eigenvector, correlate strongly with changes in unfolding times upon mutation (Fig.~\ref{fig:wt-weight-mfpt-correlation}). In particular, mutations at residues with larger values of $I_r$ tend to exhibit greater acceleration of first-passage unfolding times. For each residue $r$, we report the mean log MFPT change over a subset of single point mutations spanning diverse physicochemical properties; the full list of mutants and corresponding values is provided in Table~S1.

A natural interpretation is that $I_r$ describes how strongly residue $r$ contributes to the statistical separation between the folded and unfolded ensembles encoded by the HLDA descriptors. Residues with large $I_r$ therefore act as kinetic ``hot spots'', whereby perturbing them is more likely to disrupt interactions that support the folded basin and to promote escape, yielding faster unfolding on average. In contrast, residues with small $I_r$ contribute weakly to the folded--unfolded separation in this representation, and mutations at these sites tend to produce smaller or more variable kinetic effects, including slowing of unfolding. Importantly, the observed trend is robust to the RMSD threshold used to define first-passage unfolding events, which we denote by $t_{\mathrm{FPT}}$ (Fig.~\ref{fig:wt-weight-mfpt-correlation}(b)). The correlation increases with $t_{\mathrm{FPT}}$, peaks near $\sim 0.36$, and remains high over a broad range thereafter.

Together, these results show that a CV constructed solely from short WT trajectories already provides residue-level guidance for identifying positions where point mutations are most likely to accelerate or slow unfolding, without requiring any mutant-specific kinetic information.

\subsection*{HLDA separation correlates with unfolding kinetics across mutations}

HLDA provides a coarse approximation to the system’s reaction coordinate by identifying a one-dimensional projection that maximally separates the relevant metastable states. We hypothesized that changes in separability along such an optimized coordinate, arising from mutation-induced changes to the system itself and, consequently, to the coordinate, would reflect substantial alterations to the underlying FES of the transition. To test this hypothesis, we constructed an HLDA CV for each mutant and extracted the corresponding leading eigenvalue $\lambda$, which quantifies the folded–unfolded separation along that projection. These eigenvalues are then compared to the unfolding kinetics of the corresponding mutants.

\begin{figure*}[t]
\centering
\subfloat[]{%
  \includegraphics[width=0.49\linewidth]{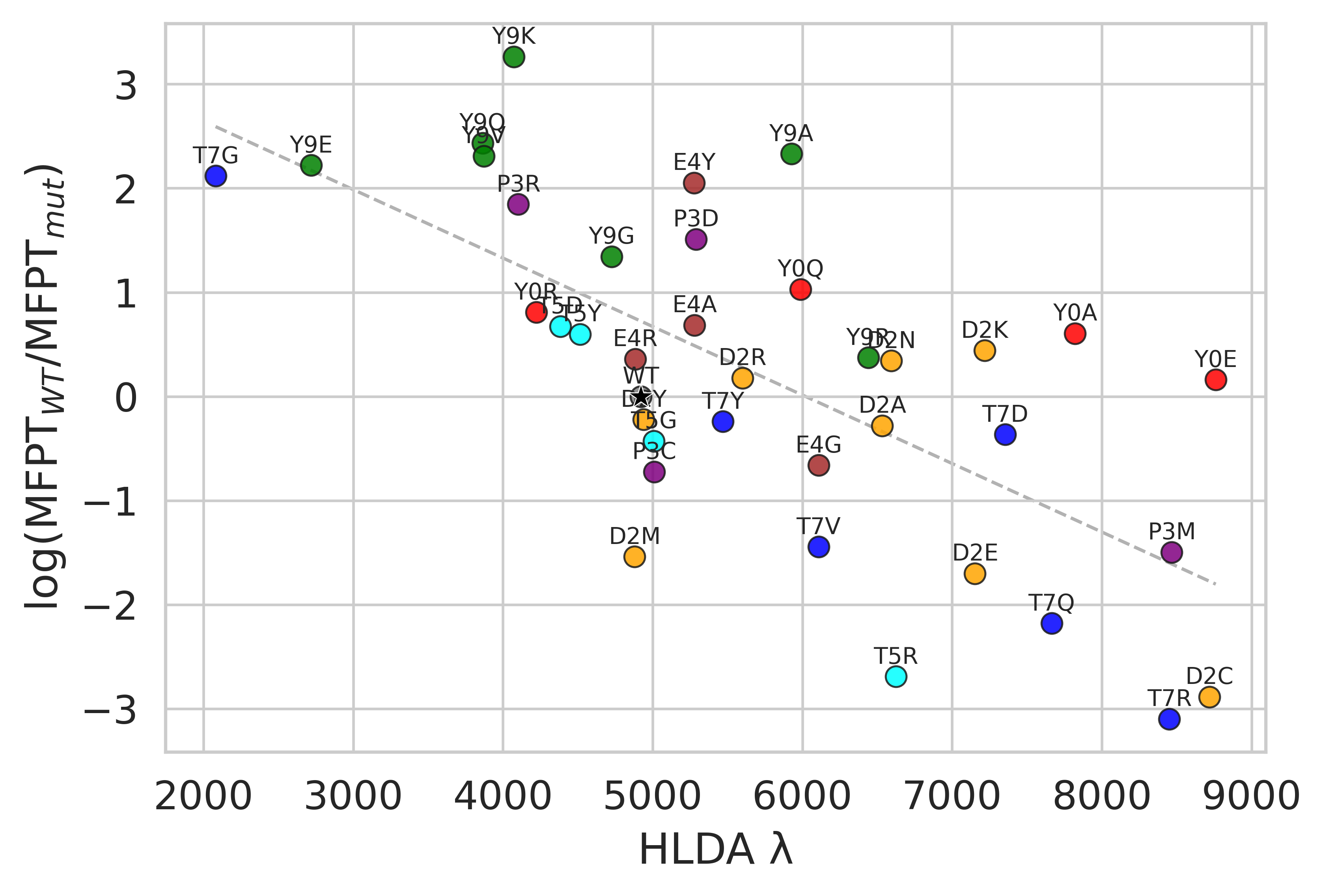}%
  \label{fig:mfpt-lambda-scatter}%
}
\hfill
\subfloat[]{%
  \includegraphics[width=0.49\linewidth]{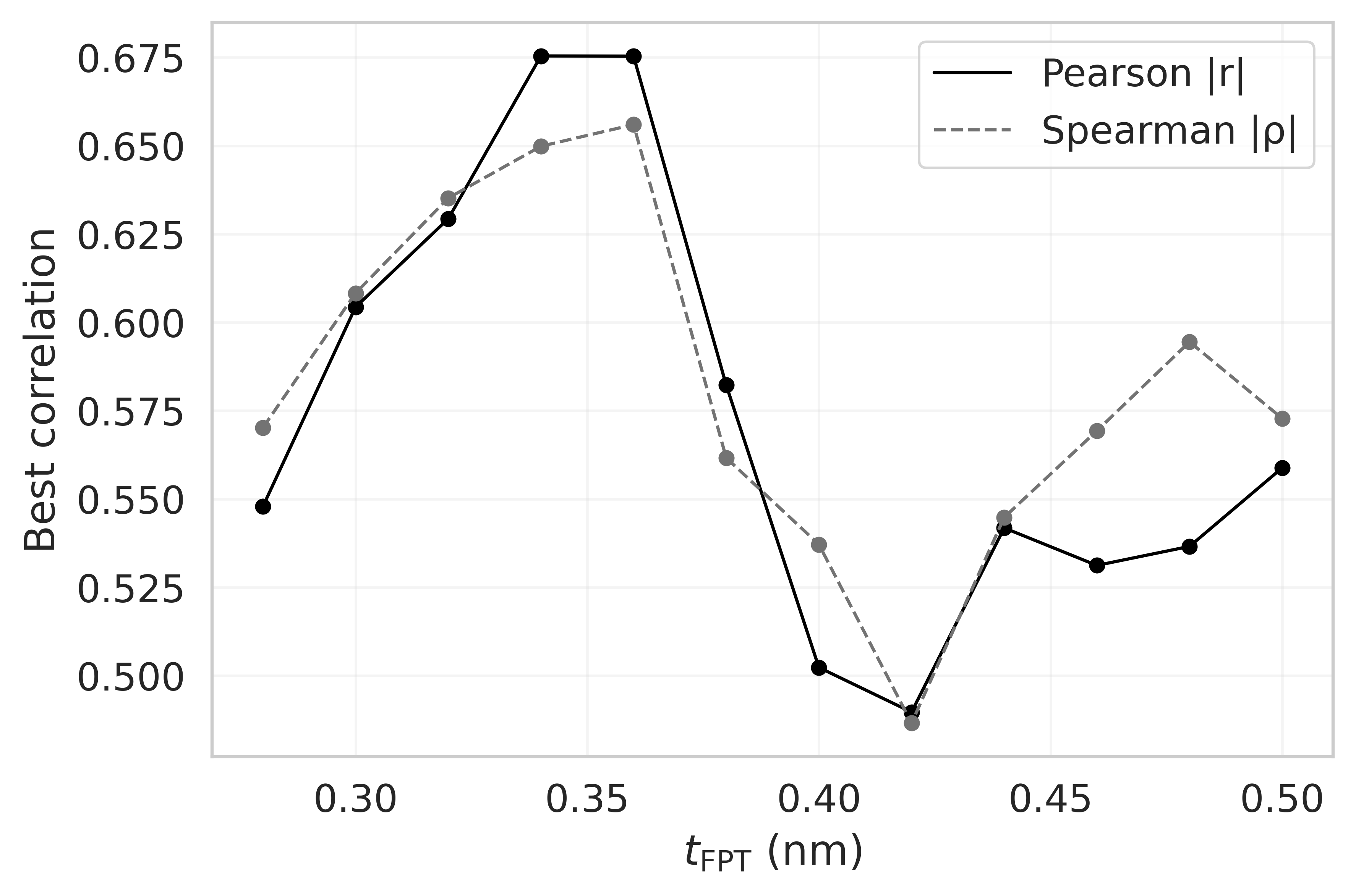}%
  \label{fig:mfpt-threshold-corr}%
}
\caption{
Relationship between HLDA eigenvalues $\lambda$ and unfolding kinetics.
(a) Logarithm of the MFPT ($t_{\mathrm{FPT}} = 0.36$) ratio for each mutant plotted against the corresponding
HLDA eigenvalue. Pearson  $r=-0.68$, $p$=$7.8\times10^{-6}$;  Spearman  $\rho=-0.66$, indicating a clear association between
state separation and unfolding kinetics.
(b) Correlation between MFPT and HLDA eigenvalue as a function of $t_{\mathrm{FPT}}$.
}
\label{fig:mfpt-hlda-correlation}
\end{figure*}

As shown in Fig.~\ref{fig:mfpt-hlda-correlation}(a), the eigenvalues $\lambda$ exhibit a clear correlation with the MFPTs across all mutations. Mutations with greater separation between the folded and unfolded ensembles, as captured by $\lambda$, systematically lead to longer MFPTs, whereas reduced separation is associated with faster transitions.

\begin{figure}[t]
\centering
\subfloat[]{%
  \includegraphics[width=0.48\linewidth]{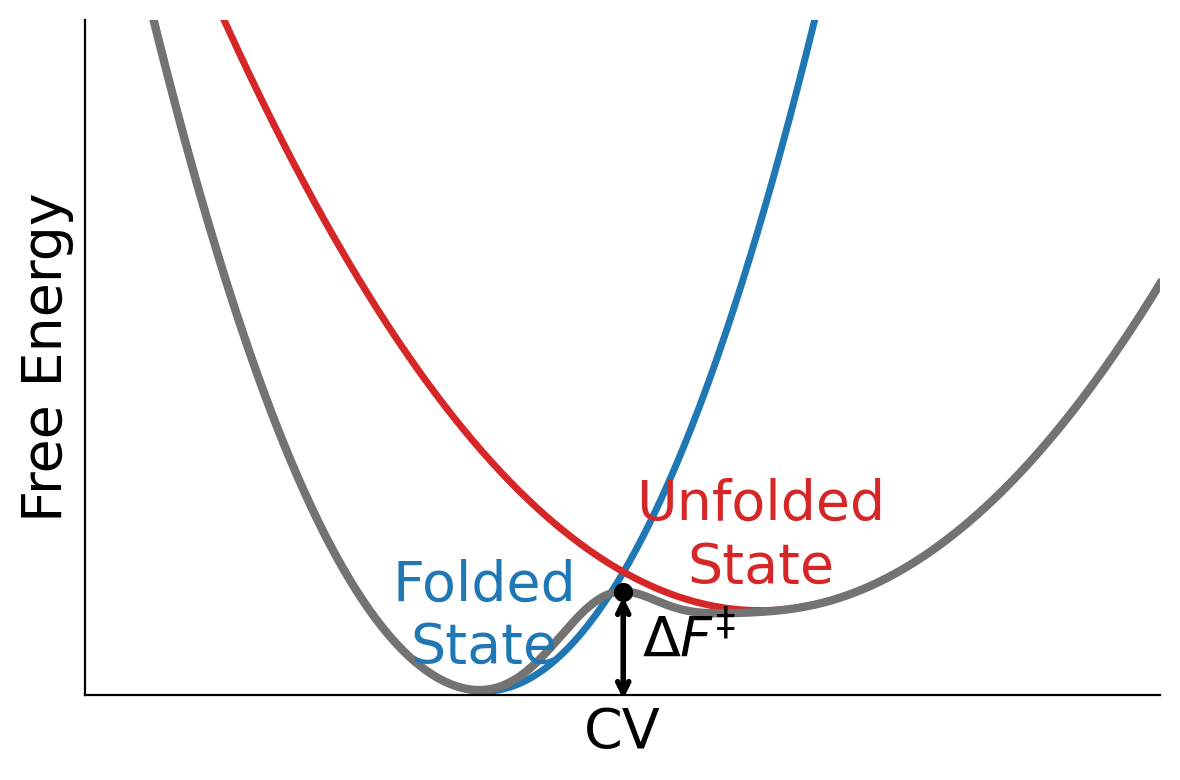}%
  \label{fig:marcus-small}%
}
\hfill
\subfloat[]{%
  \includegraphics[width=0.48\linewidth]{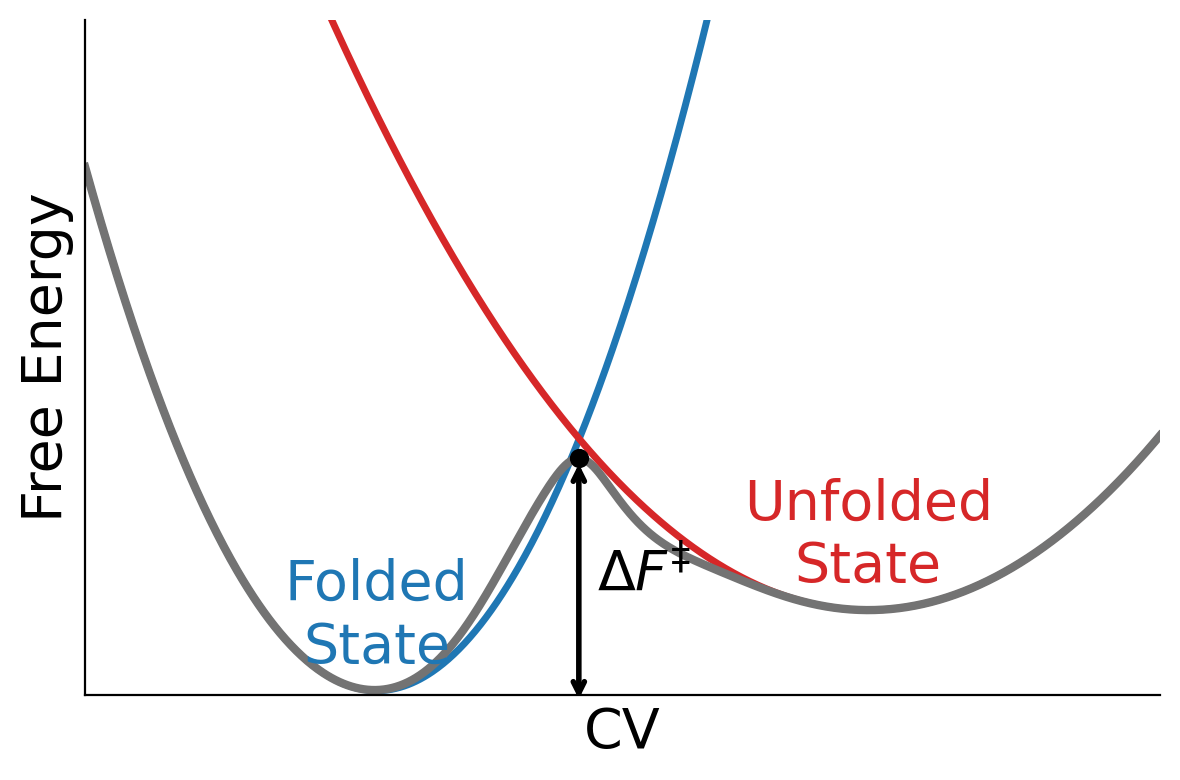}%
  \label{fig:marcus-big}%
}
\caption{
Schematic illustration of the relationship between state separation as captured by the HLDA CV and the peptide's kinetic barrier.
The folded and unfolded states are represented as parabolic free-energy basins
projected onto the one-dimensional HLDA CV.
(a) A smaller separation between the basin minima results in a lower
intersection energy and a reduced free-energy barrier.
(b) A larger separation shifts the intersection to a higher energy,
increasing the effective barrier height.
}
\label{fig:marcus-hlda}
\end{figure}

This relationship suggests a straightforward physical interpretation. A larger value of $\lambda$ is consistent with an increased effective free-energy barrier between the two basins. Geometrically, this can be understood by analogy with Marcus theory~\cite{marcus}, as illustrated in Fig.~\ref{fig:marcus-hlda}: if the free-energy landscapes of the folded and unfolded states are approximated as parabolic basins, increasing the separation between their minima raises the energy of their intersection point, thereby increasing the barrier height $\Delta F^{\ddagger}$, defined as the free-energy difference between the folded minimum and the crossing point of the two parabolas. In contrast, mutations that reduce the separation lower this intersection energy and decrease $\Delta F^{\ddagger}$, facilitating faster barrier crossing.

As in the residue-level weight analysis, the correlation between the HLDA eigenvalue and the MFPT persists across a range of $t_{\mathrm{FPT}}$ thresholds, increases with the threshold value, and reaches a maximum near 0.36 (Fig.~\ref{fig:mfpt-hlda-correlation}(b)).

\section{Discussion and Conclusions}
This work began from the recognition that tuning the free-energy surface (FES) of a peptide through point mutations is computationally prohibitive, motivating the search for a guiding light in point-mutation space. Even for Chignolin, a peptide consisting of only 10 residues, a brute-force strategy for covering its full mutation space would require evaluating MFPTs for on the order of $20^{10}$ mutants. Building on the framework of CV-FEST, the goal here was to use collective variables (CVs) to guide purposeful modifications of Chignolin's free-energy barrier through realistic chemical changes in the form of point mutations, while dramatically reducing the computational cost.

To this end, we employed Harmonic Linear Discriminant Analysis (HLDA) to construct guiding CVs. This method is attractive because it is simple to use, interpretable, and can be trained on small amounts of data originating from short unbiased simulations performed inside metastable states. A key outcome is that a CV constructed once for the wild-type (WT) system already provides residue-level guidance for identifying positions whose mutation is likely to accelerate or slow peptide unfolding. In particular, we observe a clear correlation (Fig.~\ref{fig:wt-weight-mfpt-correlation}) between the mean change in mean first-passage time (MFPT) and the WT CV weight distribution associated with each residue, supporting a practical notion of kinetic ``hot spots''. This idea is consistent with earlier work identifying functionally important regions whose perturbation modulates activity or conformational dynamics, either through data-driven analysis of key non-covalent interaction networks in protein simulations \cite{kif}, or through specific stabilizing interactions such as the Thr6–Thr8 hydrogen bond that controls the native folded state in Chignolin \cite{chignolin:mutation:induced}.

Beyond identifying residues whose mutation is expected to systematically accelerate or slow the transition, the method also provides guidance on which substitutions are likely to do so. In our framework, mutations act as controlled sequence-level perturbations of the free-energy landscape. For each mutant, we construct an HLDA CV and use the leading eigenvalue as a scalar measure of folded–unfolded separability along the one-dimensional CV axis. Across mutants, this separability shows a significant correlation with the log change in MFPTs (Fig.~\ref{fig:mfpt-hlda-correlation}), consistent with the exponential sensitivity of transition times to changes in the free-energy barrier. In this sense, the HLDA eigenvalue captures information relevant to barrier-controlled kinetics (Fig.~\ref{fig:marcus-hlda}). Taken together, MFPTs correlate with two complementary HLDA-derived quantities: a WT residue-level quantity that indicates which residues, when mutated, tend on average to accelerate or slow unfolding, and mutant-specific eigenvalues that report how strongly a given substitution differentiates the folded and unfolded ensembles. 

This correlation, while significant, is not perfect, and the remaining variance likely reflects both physical and methodological factors. The HLDA coordinate approximates the true reaction coordinate and, by construction, may miss finer features associated with the system’s slow transitions. The use of uniform folded and unfolded state definitions across mutants further contributes variability, as mutations can shift basin boundaries and barrier locations. Residual variance also reflects uncertainty in MFPT estimation due to finite sampling and the use of ST-iMetaD-based kinetic reconstruction.

A central practical takeaway is that even short trajectories sampled locally around the target conformations, without ever observing an actual transition, still carry predictive information about the transition itself, including for mutations not seen during training. In other words, MFPT-scale kinetics can be inferred, to a meaningful extent, from within-basin fluctuations alone. This distinguishes the present approach from many machine-learning-based stability optimization strategies that rely on large experimental datasets. Here, no extensive training data are required; instead, we rely solely on computationally affordable information obtained from short unbiased simulations in the metastable states of interest. At the same time, because HLDA is grounded in a physical construction, it offers the potential for mechanistic insight into how specific mutations reshape the FES, a direction left for future work.

During method development, we examined the influence of preprocessing choices, most notably the use of uniform cutoff values ($t_F$ and $t_U$) to define folded and unfolded states for HLDA training and MFPT estimation. While an optimal range of thresholds can be identified, the reported correlations persist across a relatively broad window, indicating that the results are not narrowly sensitive to a specific cutoff choice. The use of uniform thresholds therefore provides a consistent and robust baseline across mutants.

At the same time, point mutations can shift state boundaries, as reflected by changes in basin positions and barrier locations along the RMSD coordinate (Fig.~\ref{fig:ecdf-pvalue-validation}(c)). This observation further supports the idea that mutation-specific shifts in basin boundaries contribute to the residual variance in the HLDA-eigenvalue/MFPT correlation, which may arise from the use of uniform state definitions across mutants, and that tailoring thresholds per mutation could further strengthen the observed correlations. Natural extensions therefore include automated mutation-specific state identification, alternative structural descriptors beyond RMSD, and more expressive architectures for CV construction. In the near term, a pragmatic strategy is to calibrate preprocessing parameters (e.g. the state boundaries and pruning threshold) using a small validation set of mutants with full MFPT calculations, and then use these tailored definitions to efficiently explore broader mutation space. To further assess robustness and generality, future work will examine application of the approach to larger and more complex peptides and proteins.
\section{Computational Details}

\subsection*{Data Generation}
MD trajectories for each conformational state were generated by first biasing the peptide into the desired state, followed by unbiased simulations initiated from representative configurations. In the case of Chignolin, the folded and unfolded basins were initially accessed using the end-to-end distance between the terminal residues as a biasing coordinate. This distance was used solely to prepare state-restricted ensembles and was not employed in the construction of CVs or in the definition of first-passage events.

All simulations were performed on the Chignolin variant CLN025 using GROMACS patched with PLUMED, employing the CHARMM22* force field~\cite{charmm22star} and TIP3P water~\cite{tip3p}. The system was solvated in a dodecahedral box with 0.15~M NaCl, and unbiased production runs were carried out after standard equilibration.

\subsection*{Estimating Mean First Passage Time}
Following the short-time infrequent metadynamics (ST-iMetaD) approach proposed by Blumer \emph{et al.}~\cite{blumer:2024} to improve the computational efficiency of kinetic estimates in infrequent metadynamics~\cite{metadynamics:to:dynamics}, we compute the mean first-passage time (MFPT) by progressively fitting the survival probability of unfolding events to an exponential decay model. At each cutoff in the sample set, the survival function is estimated and the corresponding rate constant \(k\) is obtained together with the goodness-of-fit measure \(R^2\). The MFPT is then defined as the inverse of the rate constant associated with the statistically most reliable fit, namely the one maximizing \(R^2\).

For each mutation, 200 independent biased trajectories are generated using a bias deposition rate of \(20\,\mathrm{ns}^{-1}\). Following the protocol described in the original paper, which demonstrates that, when a well-chosen CV is employed, such as HLDA, this number of samples and bias rate are sufficient to obtain reliable kinetic estimates. Each trajectory is terminated upon reaching an RMSD of 0.5~nm. This termination criterion is used only to cap trajectory length and is chosen to exceed the range of $t_{\mathrm{FPT}}$ values considered for defining unfolding events, and is independent of the state-definition cutoffs $t_F$ and $t_U$ used for HLDA training.

To assess the validity of the obtained samples, we leverage the fact that rare-event transitions out of a long-lived metastable basin are expected to follow an exponential distribution~\cite{assessing}, with \(p(t)=k\,e^{-k t}\). Accordingly, each biased transition time is rescaled by the metadynamics acceleration factor to obtain the corresponding unbiased first-passage times \(t_i^\ast\). The exponential assumption is then tested using a Kolmogorov--Smirnov test, complemented by a Lilliefors test as recommended by Ray and Parrinello~\cite{ray:kinetics.metad}, where it is argued that the KS test alone may not be sufficient for reliable validation.

\subsection*{Tuning}
As mentioned previously, we found that the procedure shows some dependence on a small number of parameters, which requires further analysis in order to obtain a clearer picture of the resulting trends. The first parameter concerns the definition of an unfolding, or first-passage event, in terms of a structural descriptor, in particular the backbone RMSD. We denote by $t_{\mathrm{FPT}}$ the RMSD threshold used to define an unfolding event, i.e., the first time the trajectory reaches $\mathrm{RMSD} \ge t_{\mathrm{FPT}}$. Reported values for Chignolin can be found in the literature: Blumer \emph{et al.}~\cite{blumer:2024} use a value of 0.15 to indicate an unfolding, noting that this threshold is relatively low compared to those employed in OPES~\cite{Ray:OPES} or Enemark \emph{et al.}~\cite{rmsd2012}, where a value of 0.18 is used.

For this purpose, we begin by analyzing an unbiased trajectory of the WT and two point-mutants of Chignolin protein as a function of RMSD (Fig. S1). The trajectory is initiated from a minimum-enthalpy reference structure and therefore fluctuates primarily within the range \(0.15\text{–}0.35\). RMSD values around 0.35, which are comparable to thresholds used in the previously mentioned studies, may indicate partial escape from the folded basin but are not sufficient to fully overcome the free-energy barrier, as the system frequently relaxes back to the folded state.

\begin{figure}[t]
\centering
\subfloat[]{%
  \includegraphics[width=0.49\linewidth]{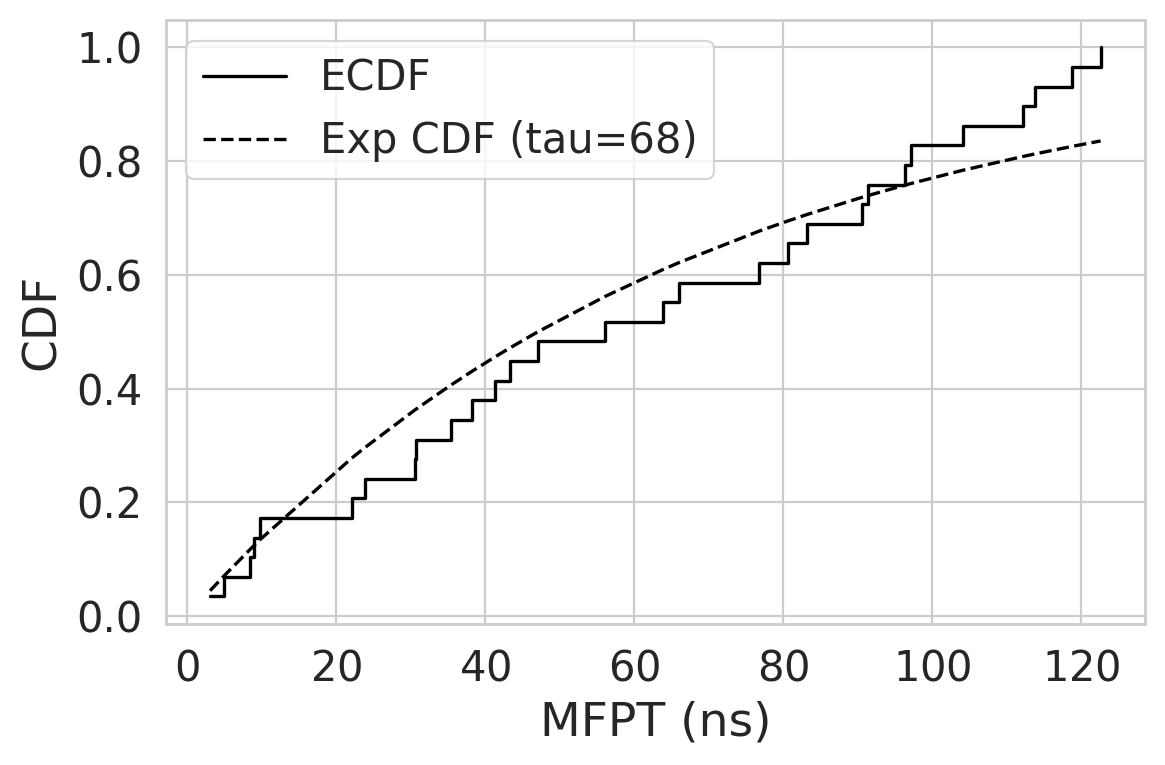}%
  \label{fig:wt-ecdf}%
}
\hfill
\subfloat[]{%
  \includegraphics[width=0.49\linewidth]{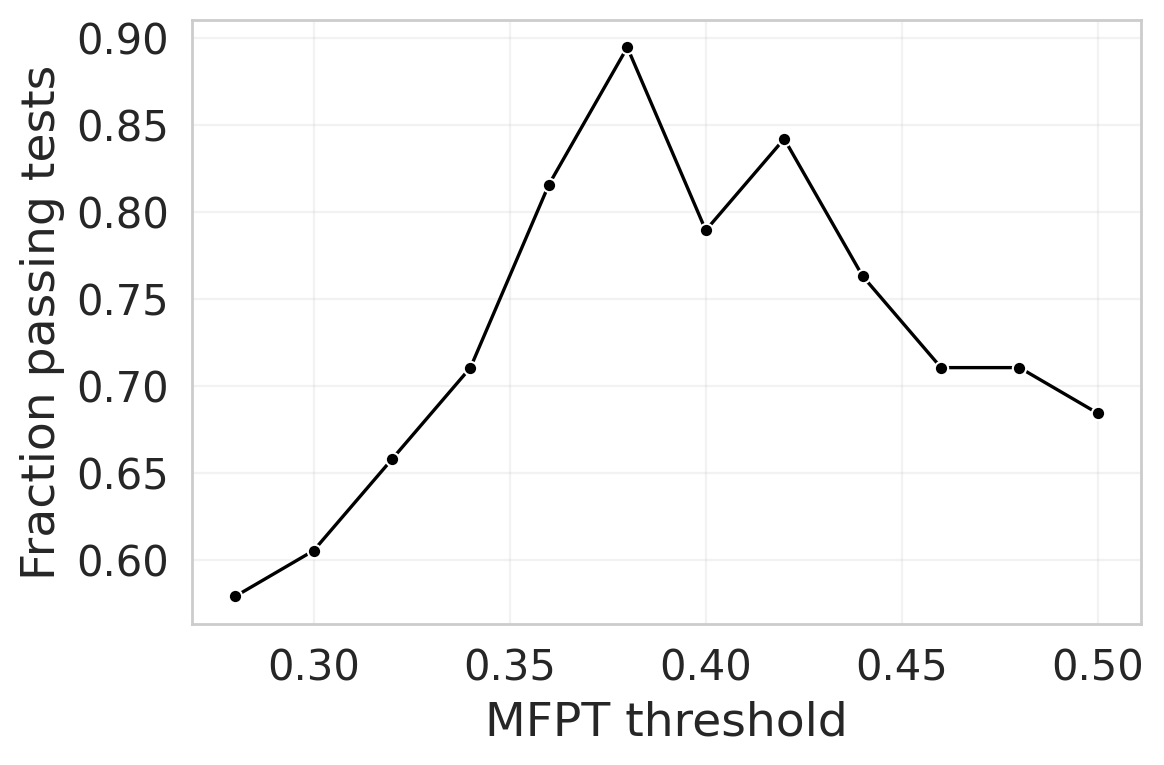}%
  \label{fig:pvalue-fraction}%
}

\vspace{0.6em}

\subfloat[]{%
  \includegraphics[width=0.7\linewidth]{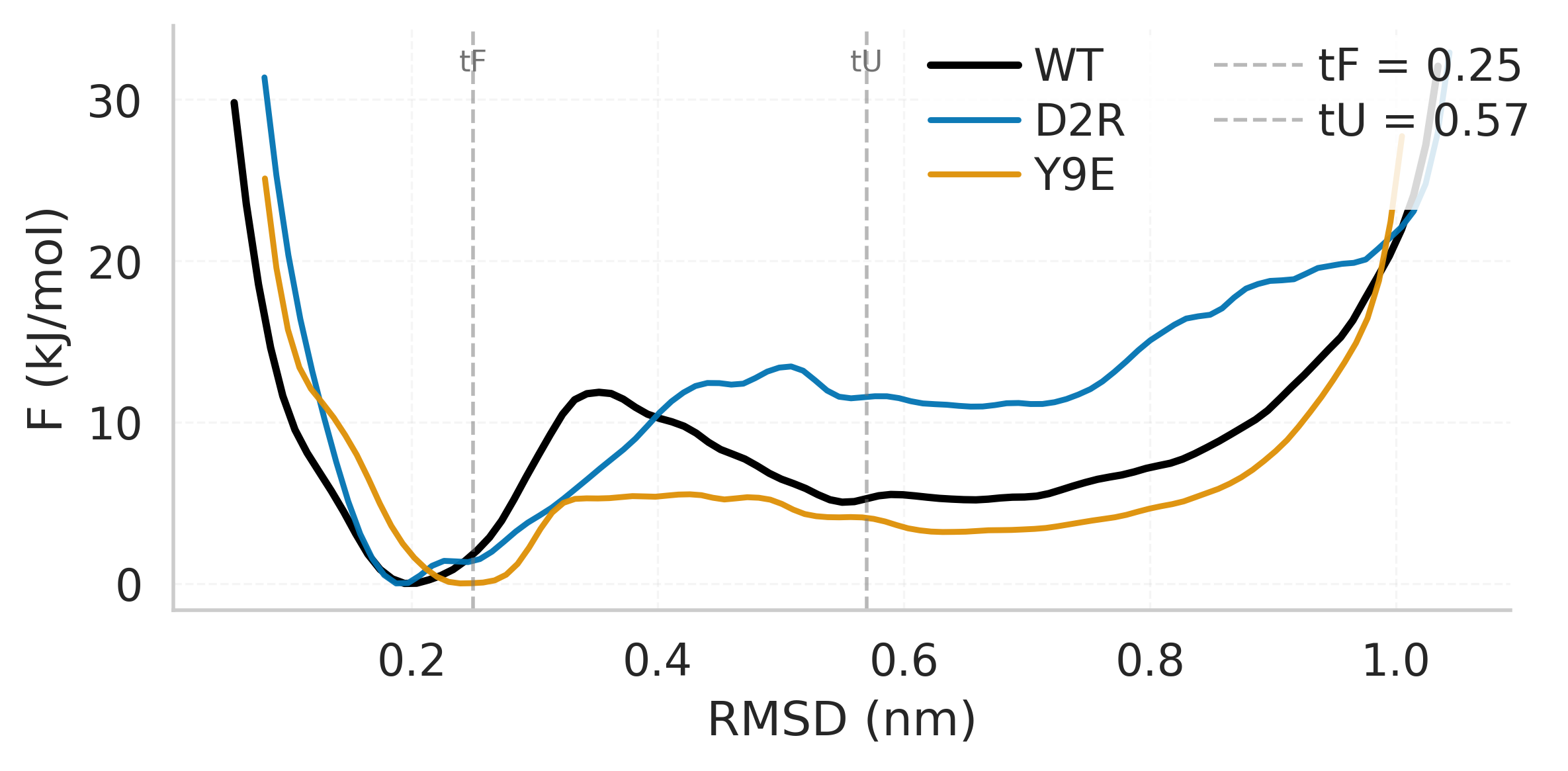}%
  \label{fig:fes}%
}

\caption{Statistical validation of the exponential first-passage-time model.
(a) Empirical cumulative distribution function (ECDF) of WT FPTs compared with the theoretical exponential CDF estimated from the fastest subset of trajectories.
(b) Fraction of mutants whose first-passage-time distributions pass the Kolmogorov--Smirnov and Lilliefors tests as a function of the RMSD threshold used to define unfolding.
(c) Free-energy surfaces (FES) as a function of RMSD for representative point mutations and the reference wild-type.}
\label{fig:ecdf-pvalue-validation}
\end{figure}

Leveraging the idea of selecting a subset of the fastest samples in order to maintain sufficiently frequent bias deposition, we fit the fastest \(1/8\) of the samples to an exponential distribution and compare the empirical cumulative distribution function to the corresponding theoretical distribution, as shown in Fig.~\ref{fig:ecdf-pvalue-validation}(a). This yields Kolmogorov--Smirnov and Lilliefors \(p\)-values of 0.33 and 0.14, respectively, indicating that the first-passage-time samples are consistent with an exponential distribution.

To ensure the robustness of the chosen RMSD threshold for detecting transitions, we applied the same fitting procedure across all mutations and scanned a range of threshold values to identify those that provide statistically consistent behavior across the full set of 36 mutants. As shown in Fig.~\ref{fig:ecdf-pvalue-validation}(b), almost all mutations satisfy the Kolmogorov--Smirnov criterion (\(p>0.2\)) and the Lilliefors criterion (\(p>0.05\)), as suggested by Ray and Parrinello~\cite{ray:kinetics.metad}.

To further validate the selected RMSD thresholds, we compare the definitions of the folded state (RMSD $\le 0.25$ nm) and unfolded state (RMSD $\ge 0.57$ nm) against the free-energy surfaces (FES) of representative mutants projected onto RMSD, as reported by Medaparambath \emph{et al.}~\cite{medaparambath2026collectivevariableguidedengineeringfreeenergy}. As shown in Fig.~\ref{fig:ecdf-pvalue-validation}(c), the vertical lines marking $t_F$ and $t_U$ bracket the barrier region separating the folded and unfolded basins, thereby excluding the barrier configurations while retaining the full folded and unfolded basins, respectively.

We now turn to the choice of parameters entering the HLDA construction, focusing in particular on the RMSD ranges used to define the folded and unfolded states. These definitions play a dual role: the selected ranges must retain sufficient structural variability to capture kinetically relevant information, while at the same time ensuring a clear separation between states and preventing cross-contamination in the HLDA training data, especially given that even unbiased trajectories may exhibit rare spontaneous partial transitions or excursions between basins. To assess the robustness of this choice, we systematically scanned a range of RMSD state boundaries and computed the resulting correlations between the HLDA eigenvalue and MFPT (Fig.~S2). Significant correlations persist across a broad window of threshold values, consistent with the location of the free-energy barriers inferred from the FES, with the region of maximal correlation reflecting an effective balance between information retention and state separation.

These observations indicate that the reported correlations are not fine-tuned to a specific threshold choice. In future work, it would be natural to extend this analysis by exploring alternative structural descriptors or CVs for state definition, as well as adopting mutation-specific, tailored threshold values to further refine the HLDA construction.

%
%

%

\section*{Supplementary Information}

Supplementary Information is available and includes additional computational details, descriptor definitions, robustness analyses, and full mutation-specific data supporting the results presented in the main text.

\begin{acknowledgments}
The authors acknowledge support from the Israel Science Foundation (ISF) under grant number 1181/24.
\end{acknowledgments}

\section*{Code availability}
All scripts and input files required to reproduce the analysis presented in this work are available at
\url{https://github.com/MendelsResearchGroup/guiding-peptide-kinetics}.

A snapshot of the repository corresponding to the version used in this study is archived at Zenodo:
\url{https://doi.org/10.5281/zenodo.18864705}.

\bibliographystyle{achemso}
\bibliography{main}

@misc{medaparambath2026collectivevariableguidedengineeringfreeenergy,
      title={Collective Variable-Guided Engineering of the Free-Energy Surface of a Small Peptide}, 
      author={Muralika Medaparambath and Alexander Zhilkin and Dan Mendels},
      year={2026},
      eprint={2602.19906},
      archivePrefix={arXiv},
      primaryClass={physics.bio-ph},
      url={https://arxiv.org/abs/2602.19906}, 
}

@article{blumer:2024,
author = {Blumer, Ofir and Reuveni, Shlomi and Hirshberg, Barak},
title = {Short-Time Infrequent Metadynamics for Improved Kinetics Inference},
journal = {Journal of Chemical Theory and Computation},
volume = {20},
number = {9},
pages = {3484-3491},
year = {2024},
doi = {10.1021/acs.jctc.4c00170},
note ={PMID: 38668722},
URL = {https://doi.org/10.1021/acs.jctc.4c00170},
eprint = { https://doi.org/10.1021/acs.jctc.4c00170}
}

@article{metadynamics:to:dynamics,
  title = {From Metadynamics to Dynamics},
  author = {Tiwary, Pratyush and Parrinello, Michele},
  journal = {Phys. Rev. Lett.},
  volume = {111},
  issue = {23},
  pages = {230602},
  numpages = {5},
  year = {2013},
  month = {Dec},
  publisher = {American Physical Society},
  doi = {10.1103/PhysRevLett.111.230602},
  url = {https://link.aps.org/doi/10.1103/PhysRevLett.111.230602}
}

@article{ligand:binding,
author = {Copeland, Robert},
year = {2015},
month = {12},
pages = {},
title = {The drug-target residence time model: A 10-year retrospective},
volume = {15},
journal = {Nature reviews. Drug discovery},
doi = {10.1038/nrd.2015.18}
}

@article{
cv-fest:dan,
author = {Dan Mendels  and Fabian Byléhn  and Timothy W. Sirk  and Juan J. de Pablo },
title = {Systematic modification of functionality in disordered elastic networks through free energy surface tailoring},
journal = {Science Advances},
volume = {9},
number = {23},
pages = {eadf7541},
year = {2023},
doi = {10.1126/sciadv.adf7541},
URL = {https://www.science.org/doi/abs/10.1126/sciadv.adf7541},
eprint = {https://www.science.org/doi/pdf/10.1126/sciadv.adf7541}}

@article{mendels:local-fluctuations,
author = {Mendels, Dan and Piccini, GiovanniMaria and Parrinello, Michele},
title = {Collective Variables from Local Fluctuations},
journal = {The Journal of Physical Chemistry Letters},
volume = {9},
number = {11},
pages = {2776-2781},
year = {2018},
doi = {10.1021/acs.jpclett.8b00733},
    note ={PMID: 29733652},
URL = { 
        https://doi.org/10.1021/acs.jpclett.8b00733
},
eprint = { 
        https://doi.org/10.1021/acs.jpclett.8b00733
}
}

@misc{hlda:dan,
      title={Collective Variables for Free Energy Surface Tailoring -- Understanding and Modifying Functionality in Systems Dominated by Rare Events}, 
      author={Dan Mendels and Juan J. de Pablo},
      year={2021},
      eprint={2108.12541},
      archivePrefix={arXiv},
      primaryClass={physics.comp-ph},
      url={https://arxiv.org/abs/2108.12541}, 
}

@article{folding-small-protein,
    author = {Mendels, Dan and Piccini, Giovannimaria and Brotzakis, Z. Faidon and Yang, Yi I. and Parrinello, Michele},
    title = {Folding a small protein using harmonic linear discriminant analysis},
    journal = {The Journal of Chemical Physics},
    volume = {149},
    number = {19},
    pages = {194113},
    year = {2018},
    month = {11},
    issn = {0021-9606},
    doi = {10.1063/1.5053566},
    url = {https://doi.org/10.1063/1.5053566},
    eprint = {https://pubs.aip.org/aip/jcp/article-pdf/doi/10.1063/1.5053566/15550067/194113\_1\_online.pdf},
}

@article{kif,
    author = {Crean, Rory M. and Slusky, Joanna S. G. and Kasson, Peter M. and Kamerlin, Shina Caroline Lynn},
    title = {KIF—Key Interactions Finder: A program to identify the key molecular interactions that regulate protein conformational changes},
    journal = {The Journal of Chemical Physics},
    volume = {158},
    number = {14},
    pages = {144114},
    year = {2023},
    month = {04},
    issn = {0021-9606},
    doi = {10.1063/5.0140882},
    url = {https://doi.org/10.1063/5.0140882},
    eprint = {https://pubs.aip.org/aip/jcp/article-pdf/doi/10.1063/5.0140882/17912972/144114\_1\_5.0140882.pdf},
}

@article{Ray:OPES,
   title={Rare Event Kinetics from Adaptive Bias Enhanced Sampling},
   volume={18},
   ISSN={1549-9626},
   url={http://dx.doi.org/10.1021/acs.jctc.2c00806},
   DOI={10.1021/acs.jctc.2c00806},
   number={11},
   journal={Journal of Chemical Theory and Computation},
   publisher={American Chemical Society (ACS)},
   author={Ray, Dhiman and Ansari, Narjes and Rizzi, Valerio and Invernizzi, Michele and Parrinello, Michele},
   year={2022},
   month=oct, pages={6500–6509} }

@Article{rmsd2012,
author ="Enemark, Søren and Rajagopalan, Raj",
title  ="Turn-directed folding dynamics of beta-hairpin-forming de novo decapeptide Chignolin",
journal  ="Phys. Chem. Chem. Phys.",
year  ="2012",
volume  ="14",
issue  ="36",
pages  ="12442-12450",
publisher  ="The Royal Society of Chemistry",
doi  ="10.1039/C2CP40285H",
url  ="http://dx.doi.org/10.1039/C2CP40285H"}

@article{assessing,
author = {Salvalaglio, Matteo and Tiwary, Pratyush and Parrinello, Michele},
title = {Assessing the Reliability of the Dynamics Reconstructed from Metadynamics},
journal = {Journal of Chemical Theory and Computation},
volume = {10},
number = {4},
pages = {1420-1425},
year = {2014},
doi = {10.1021/ct500040r},
    note ={PMID: 26580360},
URL = {https://doi.org/10.1021/ct500040r},
eprint = {https://doi.org/10.1021/ct500040r}
}

@article{
trajectory,
author = {Kresten Lindorff-Larsen  and Stefano Piana  and Ron O. Dror  and David E. Shaw },
title = {How Fast-Folding Proteins Fold},
journal = {Science},
volume = {334},
number = {6055},
pages = {517-520},
year = {2011},
doi = {10.1126/science.1208351},
URL = {https://www.science.org/doi/abs/10.1126/science.1208351},
eprint = {https://www.science.org/doi/pdf/10.1126/science.1208351}}

@article{kang2022frtpred,
title = {FRTpred: A novel approach for accurate prediction of protein folding rate and type},
journal = {Computers in Biology and Medicine},
volume = {149},
pages = {105911},
year = {2022},
issn = {0010-4825},
doi = {https://doi.org/10.1016/j.compbiomed.2022.105911},
url = {https://www.sciencedirect.com/science/article/pii/S0010482522006540},
author = {Balachandran Manavalan and Jooyoung Lee}
}

@article{prorate,
title = "Prediction of protein folding rates from structural topology and complex network properties",
author = "Jiangning Song and Kazuhiro Takemoto and Hongbin Shen and Hao Tan and Gromiha, \{M Michael\} and Tatsuya Akutsu",
year = "2010",
doi = "10.2197/ipsjtbio.3.40",
language = "English",
volume = "3",
pages = "40 -- 53",
journal = "IPSJ Transactions on Bioinformatics",
issn = "1882-6779",
}

@article{rate-prediction-review,
    author = {Chang, Catherine Ching Han and Tey, Beng Ti and Song, Jiangning and Ramanan, Ramakrishnan Nagasundara},
    title = {Towards more accurate prediction of protein folding rates: a review of the existing web-based bioinformatics approaches},
    journal = {Briefings in Bioinformatics},
    volume = {16},
    number = {2},
    pages = {314-324},
    year = {2014},
    month = {03},
    issn = {1467-5463},
    doi = {10.1093/bib/bbu007},
    url = {https://doi.org/10.1093/bib/bbu007},
    eprint = {https://academic.oup.com/bib/article-pdf/16/2/314/681193/bbu007.pdf},
}

@article{recent-protein-mech,
title = {Recent advances in the integration of protein mechanics and machine learning},
journal = {Extreme Mechanics Letters},
volume = {72},
pages = {102236},
year = {2024},
issn = {2352-4316},
doi = {https://doi.org/10.1016/j.eml.2024.102236},
url = {https://www.sciencedirect.com/science/article/pii/S2352431624001160},
author = {Yen-Lin Chen and Shu-Wei Chang}
}

@article{wang2013swfoldrate,
author = {Cheng, Xiang and Xiao, Xuan and Wu, Zhi-cheng and Wang, Pu and Lin, Wei-zhong},
title = {Swfoldrate: Predicting protein folding rates from amino acid sequence with sliding window method},
journal = {Proteins: Structure, Function, and Bioinformatics},
volume = {81},
number = {1},
pages = {140-148},
doi = {https://doi.org/10.1002/prot.24171},
url = {https://onlinelibrary.wiley.com/doi/abs/10.1002/prot.24171},
eprint = {https://onlinelibrary.wiley.com/doi/pdf/10.1002/prot.24171},
year = {2013}
}

@article{lin2006foldrate,
  author    = {Gromiha, Michael M. and Thangakani, Arul M. and Selvaraj, Samuel},
  title     = {{FOLD-RATE}: prediction of protein folding rates from amino acid sequence},
  journal   = {Nucleic Acids Research},
  year      = {2006},
  volume    = {34},
  number    = {Web Server issue},
  pages     = {W70--W74},
  doi       = {10.1093/nar/gkl043},
  pmid      = {16845101},
  pmcid     = {PMC1538837}
}

@article{2009foldrate,
author = {Chou, Kuo-Chen and Shen, Hong-Bin},
year = {2009},
month = {07},
pages = {},
title = {FoldRate: A Web-Server for Predicting Protein Folding Rates from Primary Sequence},
volume = {3},
journal = {The Open Bioinformatics Journal},
doi = {10.2174/1875036200903010031}
}

@article{zou2010seqrate,
  author  = {Zou, Hui and Lin, Guo-Neng and Wang, Zhen and Xu, Dong and Cheng, Jianlin},
  title   = {SeqRate: sequence-based protein folding type classification and rates prediction},
  journal = {BMC Bioinformatics},
  year    = {2010},
  volume  = {11},
  number  = {Suppl 3},
  pages   = {S1},
  doi     = {10.1186/1471-2105-11-S3-S1}
}

@article{wei2014predpfr,
author = {Shen, Hong-Bin and Song, Jiangning and Chou, Kuo-Chen},
year = {2009},
month = {06},
pages = {136-143},
title = {Prediction of protein folding rates from primary sequence by fusing multiple sequential features},
volume = {2},
journal = {Journal of Biomedical Science and Engineering},
doi = {10.4236/jbise.2009.23024}
}

@article{kfold,
    author = {Capriotti, E. and Casadio, R.},
    title = {K-Fold: a tool for the prediction of the protein folding kinetic order and rate},
    journal = {Bioinformatics},
    volume = {23},
    number = {3},
    pages = {385-386},
    year = {2006},
    month = {11},
    issn = {1367-4803},
    doi = {10.1093/bioinformatics/btl610},
    url = {https://doi.org/10.1093/bioinformatics/btl610},
    eprint = {https://academic.oup.com/bioinformatics/article-pdf/23/3/385/49829496/bioinformatics\_23\_3\_385.pdf},
}

@article{kpro,
title = {K-Pro: Kinetics Data on Proteins and Mutants},
journal = {Journal of Molecular Biology},
volume = {435},
number = {20},
pages = {168245},
year = {2023},
issn = {0022-2836},
doi = {https://doi.org/10.1016/j.jmb.2023.168245},
url = {https://www.sciencedirect.com/science/article/pii/S002228362300356X},
author = {Paola Turina and Piero Fariselli and Emidio Capriotti}
}

@article{kineticdb,
  title   = {KineticDB: a database of protein folding kinetics},
  author  = {Bogatyreva, Natalya S. and Osypov, Alexander A. and Ivankov, Dmitry N.},
  journal = {Nucleic Acids Research},
  year    = {2009},
  volume  = {37},
  number  = {Database issue},
  pages   = {D342--D346},
  doi     = {10.1093/nar/gkn696},
  note    = {Epub 2008-10-08}
}

@article{ray:kinetics.metad,
author = {Ray, Dhiman and Parrinello, Michele},
title = {Kinetics from Metadynamics: Principles, Applications, and Outlook},
journal = {Journal of Chemical Theory and Computation},
volume = {19},
number = {17},
pages = {5649-5670},
year = {2023},
doi = {10.1021/acs.jctc.3c00660},
note ={PMID: 37585703},
URL = { 
        https://doi.org/10.1021/acs.jctc.3c00660
},
eprint = { 
        https://doi.org/10.1021/acs.jctc.3c00660
}
}

@article{chignolin:mutation:induced,
author ="Maruyama, Yutaka and Koroku, Shunpei and Imai, Misaki and Takeuchi, Koh and Mitsutake, Ayori",
title  ="Mutation-induced change in chignolin stability from pi-turn to alpha-turn",
journal  ="RSC Adv.",
year  ="2020",
volume  ="10",
issue  ="38",
pages  ="22797-22808",
publisher  ="The Royal Society of Chemistry",
doi  ="10.1039/D0RA01148G",
url  ="http://dx.doi.org/10.1039/D0RA01148G"}

@article{granger-causaility,
author = {Sobieraj, Marcin and Setny, Piotr},
title = {Granger Causality Analysis of Chignolin Folding},
journal = {Journal of Chemical Theory and Computation},
volume = {18},
number = {3},
pages = {1936-1944},
year = {2022},
doi = {10.1021/acs.jctc.1c00945},
    note ={PMID: 35167755},
URL = { 
        https://doi.org/10.1021/acs.jctc.1c00945
},
eprint = { 
        https://doi.org/10.1021/acs.jctc.1c00945
}
}

@article{marcus,
  title = {Electron transfer reactions in chemistry. Theory and experiment},
  author = {Marcus, Rudolph A.},
  journal = {Rev. Mod. Phys.},
  volume = {65},
  issue = {3},
  pages = {599--610},
  numpages = {0},
  year = {1993},
  month = {Jul},
  publisher = {American Physical Society},
  doi = {10.1103/RevModPhys.65.599},
  url = {https://link.aps.org/doi/10.1103/RevModPhys.65.599}
}

@article{CUNNINGHAM201759,
title = {Peptides and peptidomimetics as regulators of protein–protein interactions},
journal = {Current Opinion in Structural Biology},
volume = {44},
pages = {59-66},
year = {2017},
note = {Carbohydrates: A feast of structural glycobiology • Sequences and topology: Computational studies of protein-protein interactions},
issn = {0959-440X},
doi = {https://doi.org/10.1016/j.sbi.2016.12.009},
url = {https://www.sciencedirect.com/science/article/pii/S0959440X16301300},
author = {Anna D Cunningham and Nir Qvit and Daria Mochly-Rosen}
}

@article{dynamic-personalities-of-proteins,
author = {Henzler-Wildman, Katherine and Kern, Dorothee},
year = {2008},
month = {01},
pages = {964-72},
title = {Dynamic Personalities of Proteins},
volume = {450},
journal = {Nature},
doi = {10.1038/nature06522}
}

@article{boehr-2009,
author = {Boehr, David and Nussinov, Ruth and Wright, Peter},
year = {2009},
month = {11},
pages = {789-96},
title = {The role of conformational ensembles in biomolecular recognition},
volume = {5},
journal = {Nature chemical biology},
doi = {10.1038/nchembio.232}
}

@article{keskin2016,
author = {Keskin, Ozlem and Tuncbag, Nurcan and Gursoy, Attila},
title = {Predicting Protein–Protein Interactions from the Molecular to the Proteome Level},
journal = {Chemical Reviews},
volume = {116},
number = {8},
pages = {4884-4909},
year = {2016},
doi = {10.1021/acs.chemrev.5b00683},
    note ={PMID: 27074302},
URL = { 
 https://doi.org/10.1021/acs.chemrev.5b00683
},
eprint = { 
  https://doi.org/10.1021/acs.chemrev.5b00683
}

}

@article{furman2013,
title = {Druggable protein–protein interactions – from hot spots to hot segments},
journal = {Current Opinion in Chemical Biology},
volume = {17},
number = {6},
pages = {952-959},
year = {2013},
note = {Synthetic biology • Synthetic biomolecules},
issn = {1367-5931},
doi = {https://doi.org/10.1016/j.cbpa.2013.10.011},
url = {https://www.sciencedirect.com/science/article/pii/S1367593113001798},
author = {Nir London and Barak Raveh and Ora Schueler-Furman}
}

@article{McCarty-parrinello-2017,
    author = {McCarty, James and Parrinello, Michele},
    title = {A variational conformational dynamics approach to the selection of collective variables in metadynamics},
    journal = {The Journal of Chemical Physics},
    volume = {147},
    number = {20},
    pages = {204109},
    year = {2017},
    month = {11},
    issn = {0021-9606},
    doi = {10.1063/1.4998598},
    url = {https://doi.org/10.1063/1.4998598},
    eprint = {https://pubs.aip.org/aip/jcp/article-pdf/doi/10.1063/1.4998598/13771137/204109\_1\_online.pdf},
}

@article{Invernizzi-parinello-2020,
  title = {Unified Approach to Enhanced Sampling},
  author = {Invernizzi, Michele and Piaggi, Pablo M. and Parrinello, Michele},
  journal = {Phys. Rev. X},
  volume = {10},
  issue = {4},
  pages = {041034},
  numpages = {18},
  year = {2020},
  month = {Nov},
  publisher = {American Physical Society},
  doi = {10.1103/PhysRevX.10.041034},
  url = {https://link.aps.org/doi/10.1103/PhysRevX.10.041034}
}

@article{dan:discriminant,
author = {Piccini, GiovanniMaria and Mendels, Dan and Parrinello, Michele},
title = {Metadynamics with Discriminants: A Tool for Understanding Chemistry},
journal = {Journal of Chemical Theory and Computation},
volume = {14},
number = {10},
pages = {5040-5044},
year = {2018},
doi = {10.1021/acs.jctc.8b00634},
    note ={PMID: 30222350},
URL = { 
        https://doi.org/10.1021/acs.jctc.8b00634
},
eprint = { 
        https://doi.org/10.1021/acs.jctc.8b00634
}
}

@article{dan:crystal,
    author = {Zhang, Yue-Yu and Niu, Haiyang and Piccini, GiovanniMaria and Mendels, Dan and Parrinello, Michele},
    title = {Improving collective variables: The case of crystallization},
    journal = {The Journal of Chemical Physics},
    volume = {150},
    number = {9},
    pages = {094509},
    year = {2019},
    month = {03},
    issn = {0021-9606},
    doi = {10.1063/1.5081040},
    url = {https://doi.org/10.1063/1.5081040},
    eprint = {https://pubs.aip.org/aip/jcp/article-pdf/doi/10.1063/1.5081040/15556242/094509\_1\_online.pdf},
}

@article{dan:blind,
author = {Rizzi, Valerio and Mendels, Dan and Sicilia, Emilia and Parrinello, Michele},
title = {Blind Search for Complex Chemical Pathways Using Harmonic Linear Discriminant Analysis},
journal = {Journal of Chemical Theory and Computation},
volume = {15},
number = {8},
pages = {4507-4515},
year = {2019},
doi = {10.1021/acs.jctc.9b00358},
    note ={PMID: 31314521},
URL = { 
        https://doi.org/10.1021/acs.jctc.9b00358
},
eprint = { 
        https://doi.org/10.1021/acs.jctc.9b00358
}
}

@article{tip3p,
    author = {Jorgensen, William L. and Chandrasekhar, Jayaraman and Madura, Jeffry D. and Impey, Roger W. and Klein, Michael L.},
    title = {Comparison of simple potential functions for simulating liquid water},
    journal = {The Journal of Chemical Physics},
    volume = {79},
    number = {2},
    pages = {926-935},
    year = {1983},
    month = {07},
    issn = {0021-9606},
    doi = {10.1063/1.445869},
    url = {https://doi.org/10.1063/1.445869},
    eprint = {https://pubs.aip.org/aip/jcp/article-pdf/79/2/926/18942849/926\_1\_online.pdf},
}

@article{charmm22star,
title = {How Robust Are Protein Folding Simulations with Respect to Force Field Parameterization?},
journal = {Biophysical Journal},
volume = {100},
number = {9},
pages = {L47-L49},
year = {2011},
issn = {0006-3495},
doi = {https://doi.org/10.1016/j.bpj.2011.03.051},
url = {https://www.sciencedirect.com/science/article/pii/S0006349511004097},
author = {Stefano Piana and Kresten Lindorff-Larsen and DavidE. Shaw}
}

@article{salman-data-scar,
    author = {Salman, Salman N. and Shteingolts, Sergey A. and Levie, Ron and Mendels, Dan},
    title = {Evaluating the use of a machine learning simulator for structure–property prediction: A case study on disordered elastic networks},
    journal = {The Journal of Chemical Physics},
    volume = {163},
    number = {12},
    pages = {124115},
    year = {2025},
    month = {09},
    issn = {0021-9606},
    doi = {10.1063/5.0282871},
    url = {https://doi.org/10.1063/5.0282871},
    eprint = {https://pubs.aip.org/aip/jcp/article-pdf/doi/10.1063/5.0282871/20712044/124115_1_5.0282871.pdf},
}

\clearpage
\FloatBarrier

\onecolumngrid

\section{Supplementary Information}
\begin{figure}
    \centering
    \includegraphics[width=0.8\linewidth]{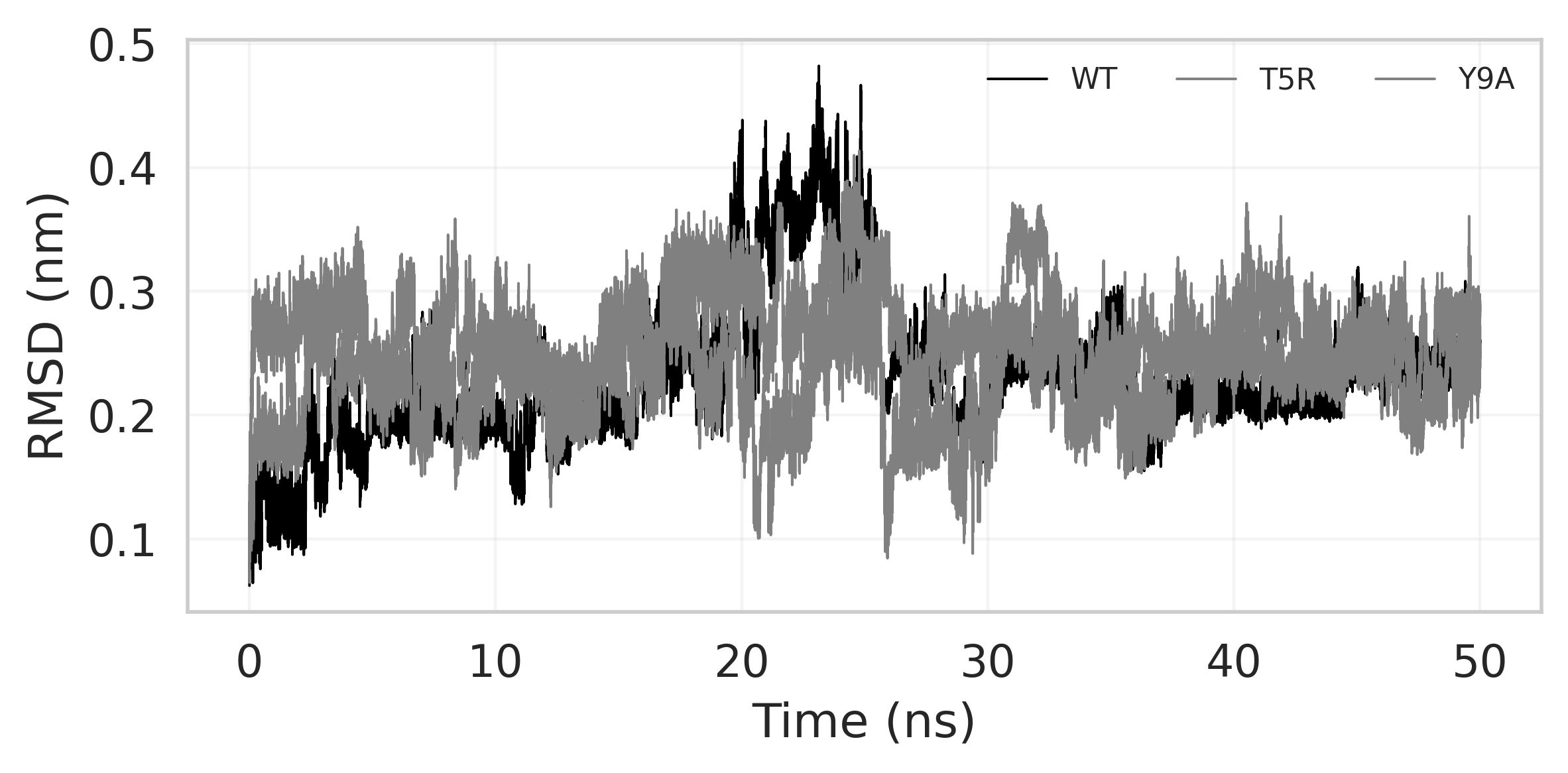}
    \caption{Chignolin WT and two point-mutations backbone RMSD from a reference folded structure corresponding to the enthalpic minimum of the native hairpin, computed along an unbiased short trajectory.}
    \label{fig:wt:rmsd}
\end{figure}

\begin{table*}[t]
\centering
\caption{Mutant-specific unfolding kinetics and HLDA separability for all simulated Chignolin variants at $t_{\mathrm{FPT}}=0.34$~nm, with folded and unfolded state definitions given by $t_F=0.25$~nm and $t_U=0.57$~nm, respectively.}
\label{tab:mutant-mfpt-lambda-1}
\begin{tabular}{l c c c}
\hline
\textbf{Mutant} & \textbf{MFPT ($\mu$s)} & \textbf{$\log(\mathrm{MFPT}_{\mathrm{WT}}/\mathrm{MFPT}_{\mathrm{mut}})$} & \textbf{HLDA eigenvalue $\lambda$} \\
\hline
WT & 0.625292 & 0.000000 & 4919.89 \\
Y0A (Tyr$\rightarrow$Ala) & 0.404168 & 0.436389 & 7821.38 \\
Y0E (Tyr$\rightarrow$Glu) & 0.309572 & 0.703029 & 8761.26 \\
Y0Q (Tyr$\rightarrow$Gln) & 0.154364 & 1.398910 & 5988.56 \\
Y0R (Tyr$\rightarrow$Arg) & 0.231474 & 0.993750 & 4223.07 \\
D2A (Asp$\rightarrow$Ala) & 0.819860 & -0.270915 & 6533.15 \\
D2C (Asp$\rightarrow$Cys) & 8.187270 & -2.572120 & 8717.71 \\
D2E (Asp$\rightarrow$Glu) & 2.873120 & -1.524940 & 7152.95 \\
D2K (Asp$\rightarrow$Lys) & 0.285676 & 0.783361 & 7217.19 \\
D2M (Asp$\rightarrow$Met) & 1.977810 & -1.151530 & 4879.15 \\
D2N (Asp$\rightarrow$Asn) & 0.395252 & 0.458695 & 6592.66 \\
D2R (Asp$\rightarrow$Arg) & 0.130910 & 1.563710 & 5599.75 \\
D2Y (Asp$\rightarrow$Tyr) & 0.673443 & -0.074184 & 4939.27 \\
P3C (Pro$\rightarrow$Cys) & 0.858174 & -0.316588 & 5011.25 \\
P3D (Pro$\rightarrow$Asp) & 0.0681503 & 2.216500 & 5290.57 \\
P3M (Pro$\rightarrow$Met) & 2.623820 & -1.434170 & 8467.28 \\
P3R (Pro$\rightarrow$Arg) & 0.0600011 & 2.343860 & 4102.90 \\
E4A (Glu$\rightarrow$Ala) & 0.164000 & 1.338350 & 5279.98 \\
E4G (Glu$\rightarrow$Gly) & 1.221300 & -0.669453 & 6107.72 \\
E4R (Glu$\rightarrow$Arg) & 0.507556 & 0.208612 & 4883.88 \\
\hline
\end{tabular}
\end{table*}

\begin{table*}[t]
\centering
\caption{Table~\ref{tab:mutant-mfpt-lambda-1}, continued.}
\label{tab:mutant-mfpt-lambda-2}
\begin{tabular}{l c c c}
\hline
\textbf{Mutant} & \textbf{MFPT ($\mu$s)} & \textbf{$\log(\mathrm{MFPT}_{\mathrm{WT}}/\mathrm{MFPT}_{\mathrm{mut}})$} & \textbf{HLDA eigenvalue $\lambda$} \\
\hline
E4Y (Glu$\rightarrow$Tyr) & 0.0610672 & 2.326240 & 5277.81 \\
T5D (Thr$\rightarrow$Asp) & 0.284876 & 0.786166 & 4382.62 \\
T5G (Thr$\rightarrow$Gly) & 1.213010 & -0.662639 & 5007.47 \\
T5R (Thr$\rightarrow$Arg) & 9.901650 & -2.762240 & 6624.71 \\
T5Y (Thr$\rightarrow$Tyr) & 0.0888015 & 1.951810 & 4515.27 \\
T7D (Thr$\rightarrow$Asp) & 0.754653 & -0.188039 & 7353.58 \\
T7G (Thr$\rightarrow$Gly) & 0.0174231 & 3.580420 & 2080.98 \\
T7Q (Thr$\rightarrow$Gln) & 1.115220 & -0.578592 & 7664.95 \\
T7R (Thr$\rightarrow$Arg) & 15.299200 & -3.197340 & 8451.40 \\
T7V (Thr$\rightarrow$Val) & 3.076060 & -1.593190 & 6107.72 \\
T7Y (Thr$\rightarrow$Tyr) & 0.290318 & 0.767243 & 5469.91 \\
Y9A (Tyr$\rightarrow$Ala) & 0.0370596 & 2.825690 & 5926.13 \\
Y9E (Tyr$\rightarrow$Glu) & 0.0444394 & 2.644090 & 2720.06 \\
Y9G (Tyr$\rightarrow$Gly) & 0.0410230 & 2.724090 & 4726.74 \\
Y9K (Tyr$\rightarrow$Lys) & 0.0165800 & 3.630020 & 4072.33 \\
Y9Q (Tyr$\rightarrow$Gln) & 0.0360310 & 2.853840 & 3864.45 \\
Y9R (Tyr$\rightarrow$Arg) & 0.3163610 & 0.681334 & 6439.89 \\
Y9V (Tyr$\rightarrow$Val) & 0.0693590 & 2.198920 & 3872.36 \\
\hline
\end{tabular}
\end{table*}
\begin{sidewaystable}
\caption{HLDA descriptor-wise contributions for Chignolin point mutations. Descriptors (e.g., d03) denote backbone distances between residues 0 and 3 (0-indexed). Blank entries indicate descriptors pruned during preprocessing due to Spearman correlation ($\rho > 0.93$).}
\centering
\scriptsize 
\begin{adjustbox}{max width=1.01\textwidth}
\begin{tabular}{lrrrrrrrrrrrrrrrrrrrrrrrrrrrr}
\toprule
descriptor & d03 & d04 & d05 & d06 & d07 & d08 & d09 & d14 & d15 & d16 & d17 & d18 & d19 & d25 & d26 & d27 & d28 & d29 & d36 & d37 & d38 & d39 & d47 & d48 & d49 & d58 & d59 & d69 \\
system &  &  &  &  &  &  &  &  &  &  &  &  &  &  &  &  &  &  &  &  &  &  &  &  &  &  &  &  \\
\midrule
AYDPETGTWY & 0.25 & -0.42 & 0.38 & -0.2 & - & -0.011 & -0.034 & 0.028 & - & - & -0.036 & - & -0.0055 & -0.086 & -0.063 & 0.086 & 0.063 & -0.21 & 0.1 & -0.06 & -0.085 & -0.082 & -0.057 & -0.051 & 0.53 & 0.074 & -0.4 & 0.19 \\
EYDPETGTWY & 0.25 & -0.4 & 0.33 & -0.17 & -0.032 & -0.022 & 0.0029 & 0.11 & -0.12 & -0.053 & -0.17 & 0.22 & -0.098 & -0.12 & -0.017 & 0.23 & -0.26 & 0.012 & 0.1 & 0.015 & -0.14 & -0.13 & -0.18 & 0.1 & 0.37 & 0.13 & -0.36 & 0.18 \\
QYDPETGTWY & 0.17 & -0.36 & 0.34 & -0.1 & -0.049 & -0.025 & -0.0085 & 0.072 & 0.058 & -0.037 & -0.21 & 0.22 & -0.083 & -0.17 & -0.023 & 0.32 & -0.33 & -0.012 & 0.068 & 0.015 & -0.12 & -0.053 & -0.24 & 0.22 & 0.33 & 0.066 & -0.32 & 0.11 \\
RYDPETGTWY & 0.098 & -0.36 & 0.4 & -0.078 & -0.079 & -0.023 & -0.023 & 0.087 & 0.032 & -0.083 & -0.15 & 0.22 & -0.11 & -0.18 & 0.0073 & 0.3 & -0.37 & 0.026 & 0.052 & -0.023 & -0.097 & 0.028 & -0.22 & 0.23 & 0.29 & 0.064 & -0.35 & 0.092 \\
YYAPETGTWY & 0.21 & -0.42 & 0.42 & -0.2 & - & - & -0.056 & - & - & - & - & -0.029 & - & -0.12 & 0.0032 & 0.069 & - & -0.21 & 0.051 & -0.035 & -0.085 & 0.0067 & -0.11 & 0.047 & 0.52 & 0.086 & -0.42 & 0.14 \\
YYCPETGTWY & - & -0.11 & -0.054 & 0.18 & - & - & 0.082 & - & - & - & - & - & 0.077 & 0.26 & 0.093 & 0.03 & 0.021 & 0.39 & -0.074 & 0.17 & 0.37 & -0.66 & 0.073 & -0.27 & 0.17 & - & - & - \\
YYDCETGTWY & 0.14 & 0.16 & -0.25 & -0.2 & 0.088 & 0.12 & 0.041 & -0.097 & -0.15 & 0.22 & 0.15 & -0.31 & 0.19 & 0.24 & 0.0019 & -0.29 & 0.3 & 0.054 & -0.042 & 0.027 & 0.21 & -0.4 & 0.19 & -0.26 & -0.038 & 0.0059 & 0.18 & 0.066 \\
YYDDETGTWY & 0.19 & -0.38 & 0.31 & -0.084 & -0.036 & -0.061 & 0.0045 & 0.12 & 0.044 & 0.021 & -0.26 & 0.22 & -0.065 & -0.2 & -0.047 & 0.28 & -0.24 & -0.13 & - & 0.17 & -0.23 & 0.0064 & -0.3 & 0.3 & 0.27 & -0.028 & -0.19 & 0.098 \\
YYDMETGTWY & 0.17 & -0.39 & 0.34 & -0.078 & -0.027 & -0.049 & -0.0033 & 0.11 & 0.026 & -0.0063 & -0.23 & 0.21 & -0.048 & -0.17 & -0.096 & 0.29 & -0.16 & -0.17 & 0.089 & 0.082 & -0.3 & 0.063 & -0.28 & 0.29 & 0.27 & -0.0076 & -0.23 & 0.11 \\
YYDPATGTWY & 0.26 & -0.49 & 0.32 & - & - & -0.17 & 0.047 & 0.15 & -0.031 & -0.14 & -0.1 & 0.25 & -0.14 & -0.056 & -0.074 & 0.12 & -0.044 & -0.12 & 0.13 & -0.0022 & -0.22 & -0.053 & -0.084 & 0.017 & 0.43 & 0.055 & -0.31 & 0.14 \\
YYDPEDGTWY & 0.31 & -0.41 & 0.27 & -0.15 & - & -0.054 & 0.022 & 0.037 & - & - & -0.049 & - & -0.016 & -0.058 & -0.13 & 0.088 & 0.17 & -0.29 & 0.13 & -0.069 & -0.12 & -0.12 & -0.024 & -0.13 & 0.55 & 0.048 & -0.3 & 0.18 \\
YYDPEGGTWY & 0.43 & -0.34 & 0.2 & -0.3 & - & - & - & -0.0071 & - & - & - & - & - & -0.075 & -0.071 & 0.041 & - & -0.23 & 0.14 & -0.083 & -0.071 & -0.21 & -0.039 & -0.082 & 0.5 & 0.12 & -0.31 & 0.25 \\
YYDPERGTWY & 0.32 & -0.41 & 0.31 & -0.23 & - & - & - & 0.029 & - & - & - & - & - & -0.096 & -0.021 & - & - & -0.21 & 0.075 & -0.041 & -0.028 & -0.16 & -0.013 & -0.12 & 0.55 & 0.12 & -0.37 & 0.19 \\
YYDPETGDWY & 0.33 & -0.37 & 0.33 & -0.32 & - & 0.021 & -0.035 & 0.033 & - & - & -0.025 & - & -0.029 & -0.097 & 0.074 & -0.04 & - & -0.094 & - & 0.089 & -0.059 & -0.21 & -0.031 & -0.1 & 0.47 & 0.12 & -0.39 & 0.26 \\
YYDPETGGWY & -0.18 & 0.43 & -0.32 & - & - & 0.16 & -0.023 & -0.083 & -0.013 & 0.15 & 0.15 & -0.31 & 0.23 & 0.044 & -0.016 & -0.14 & 0.29 & -0.041 & -0.063 & 0.0023 & 0.12 & -0.0009 & 0.077 & -0.12 & -0.38 & -0.048 & 0.39 & -0.073 \\
YYDPETGQWY & 0.29 & -0.48 & 0.37 & -0.12 & -0.061 & -0.051 & 0.017 & 0.12 & -0.078 & 0.0044 & -0.094 & 0.12 & -0.045 & -0.049 & -0.14 & 0.2 & -0.068 & -0.14 & 0.12 & -0.0091 & -0.17 & -0.12 & -0.14 & 0.11 & 0.42 & 0.023 & -0.31 & 0.19 \\
YYDPETGRWY & -0.34 & 0.41 & -0.32 & 0.25 & - & - & - & - & - & - & - & - & - & 0.047 & 0.0057 & - & - & 0.12 & -0.068 & 0.0029 & 0.021 & 0.24 & - & 0.078 & -0.52 & -0.089 & 0.38 & -0.23 \\
YYDPETGTWA & - & - & 0.046 & - & - & -0.089 & -0.16 & - & - & - & -0.2 & -0.026 & - & -0.22 & -0.22 & 0.0013 & - & -0.55 & -0.09 & 0.1 & -0.32 & 0.49 & -0.035 & 0.0072 & 0.34 & - & -0.2 & - \\
YYDPETGTWE & 0.23 & -0.36 & 0.34 & -0.084 & -0.15 & 0.0055 & 0.028 & - & - & - & -0.16 & 0.085 & -0.058 & -0.18 & -0.12 & 0.36 & -0.27 & -0.075 & 0.051 & -0.074 & -0.0065 & -0.14 & -0.22 & 0.18 & 0.42 & -0.011 & -0.28 & 0.12 \\
YYDPETGTWG & 0.23 & -0.32 & 0.29 & - & -0.24 & 0.017 & - & - & - & - & -0.1 & - & - & -0.18 & -0.16 & 0.4 & -0.3 & -0.083 & 0.051 & -0.11 & 0.089 & -0.22 & -0.23 & 0.16 & 0.42 & -0.022 & -0.21 & 0.054 \\
YYDPETGTWK & 0.26 & -0.34 & 0.16 & - & - & -0.13 & 0.026 & -0.0027 & - & - & -0.076 & - & -0.031 & 0.025 & -0.36 & 0.12 & 0.23 & -0.44 & 0.21 & -0.15 & -0.27 & 0.12 & -0.0056 & 0.07 & 0.38 & -0.099 & -0.18 & 0.16 \\
YYDPETGTWQ & - & 0.036 & -0.029 & 0.029 & - & 0.063 & -0.013 & - & -0.15 & 0.083 & 0.25 & -0.3 & 0.21 & 0.12 & -0.0058 & -0.25 & 0.46 & -0.15 & - & -0.13 & 0.29 & -0.2 & 0.26 & -0.47 & 0.12 & - & 0.11 & -0.046 \\
YYDPETGTWR & - & - & 0.15 & -0.16 & -0.43 & 0.036 & 0.0013 & - & -0.065 & 0.01 & 0.11 & -0.13 & -0.24 & -0.11 & -0.2 & 0.022 & 0.0078 & -0.34 & - & - & -0.34 & 0.56 & - & - & - & - & -0.16 & 0.22 \\
YYDPETGTWV & - & 0.2 & -0.34 & 0.06 & 0.13 & -0.024 & 0.088 & - & -0.15 & 0.23 & 0.046 & -0.22 & 0.2 & 0.16 & -0.1 & -0.16 & 0.42 & -0.17 & -0.054 & 0.061 & 0.11 & -0.18 & 0.076 & -0.18 & -0.17 & -0.11 & 0.48 & -0.12 \\
YYDPETGVWY & 0.25 & -0.15 & 0.1 & -0.24 & - & - & - & - & - & 0.17 & -0.18 & - & - & -0.27 & 0.06 & - & - & -0.51 & -0.17 & 0.29 & -0.31 & 0.31 & -0.23 & 0.24 & 0.18 & - & - & - \\
YYDPETGYWY & 0.1 & -0.016 & 0.078 & -0.24 & - & 0.2 & -0.037 & 0.014 & -0.1 & 0.1 & 0.13 & -0.22 & 0.23 & 0.053 & 0.13 & -0.18 & - & 0.31 & -0.066 & 0.064 & 0.32 & -0.57 & 0.12 & -0.29 & 0.12 & 0.15 & -0.11 & 0.12 \\
YYDPEYGTWY & 0.26 & -0.39 & 0.25 & - & - & -0.17 & - & -0.096 & - & - & - & - & - & 0.0016 & -0.32 & 0.14 & - & -0.31 & 0.14 & -0.19 & -0.084 & 0.02 & -0.04 & 0.19 & 0.49 & -0.17 & -0.29 & 0.077 \\
YYDPGTGTWY & 0.25 & -0.47 & 0.41 & -0.14 & -0.01 & -0.092 & 0.0083 & 0.12 & -0.056 & -0.055 & -0.11 & 0.19 & -0.099 & -0.065 & -0.041 & 0.15 & -0.11 & -0.075 & 0.11 & -0.025 & -0.13 & -0.082 & -0.091 & 0.036 & 0.43 & 0.065 & -0.37 & 0.17 \\
YYDPRTGTWY & 0.32 & -0.43 & 0.36 & -0.24 & - & -0.027 & -0.0047 & 0.047 & - & - & -0.056 & - & -0.023 & -0.13 & 0.042 & 0.049 & - & -0.17 & - & 0.035 & -0.0019 & -0.18 & -0.074 & -0.082 & 0.52 & 0.075 & -0.34 & 0.18 \\
YYDPYTGTWY & 0.29 & -0.062 & -0.15 & - & - & - & - & -0.17 & - & - & - & - & - & -0.016 & -0.34 & 0.2 & - & -0.5 & 0.24 & -0.2 & -0.27 & 0.16 & -0.23 & 0.17 & 0.38 & 0.12 & -0.11 & 0.092 \\
YYDRETGTWY & 0.14 & -0.38 & 0.32 & -0.051 & - & -0.077 & -0.0086 & 0.14 & 0.031 & -0.051 & -0.24 & 0.27 & -0.086 & -0.15 & -0.11 & 0.3 & -0.14 & -0.17 & 0.13 & 0.03 & -0.33 & 0.13 & -0.23 & 0.22 & 0.27 & 0.046 & -0.26 & 0.11 \\
YYEPETGTWY & 0.34 & -0.39 & 0.31 & -0.25 & - & - & - & - & - & - & - & - & - & -0.061 & 0.0096 & - & - & -0.11 & 0.066 & 0.005 & -0.017 & -0.25 & -0.021 & -0.091 & 0.52 & 0.1 & -0.38 & 0.23 \\
YYKPETGTWY & 0.26 & -0.41 & 0.24 & - & - & -0.16 & 0.071 & 0.11 & - & - & -0.26 & 0.24 & -0.057 & -0.1 & -0.17 & 0.34 & -0.091 & -0.2 & 0.13 & 0.036 & -0.24 & -0.078 & -0.27 & 0.23 & 0.32 & -0.077 & -0.099 & 0.081 \\
YYMPETGTWY & - & - & 0.024 & - & - & 0.095 & -0.0049 & - & - & - & 0.24 & -0.27 & 0.19 & 0.056 & 0.12 & -0.28 & 0.25 & 0.17 & -0.069 & -0.012 & 0.42 & -0.44 & 0.27 & -0.43 & 0.095 & - & - & - \\
YYNPETGTWY & - & - & 0.3 & - & -0.48 & 0.17 & 0.0033 & - & -0.13 & 0.031 & 0.13 & -0.19 & -0.0082 & -0.22 & -0.21 & 0.39 & -0.33 & 0.0077 & 0.092 & -0.37 & 0.27 & -0.017 & -0.06 & - & 0.023 & - & -0.02 & 0.011 \\
YYRPETGTWY & - & - & 0.046 & - & - & -0.089 & -0.16 & - & - & - & -0.2 & -0.026 & - & -0.22 & -0.22 & 0.0013 & - & -0.55 & -0.09 & 0.1 & -0.32 & 0.49 & -0.035 & 0.0072 & 0.34 & - & -0.2 & - \\
YYYPETGTWY & 0.24 & -0.39 & 0.28 & - & - & -0.17 & -0.052 & -0.061 & - & - & -0.11 & - & -0.076 & -0.067 & -0.18 & 0.23 & - & -0.32 & 0.097 & -0.16 & 0.0014 & -0.03 & -0.12 & -0.0023 & 0.59 & -0.048 & -0.25 & 0.021 \\
\bottomrule

\end{tabular}
\end{adjustbox}

\end{sidewaystable}

\begin{figure}[t]
    \centering
    \makebox[\textwidth][c]{%
        \includegraphics[width=1.08\textwidth]{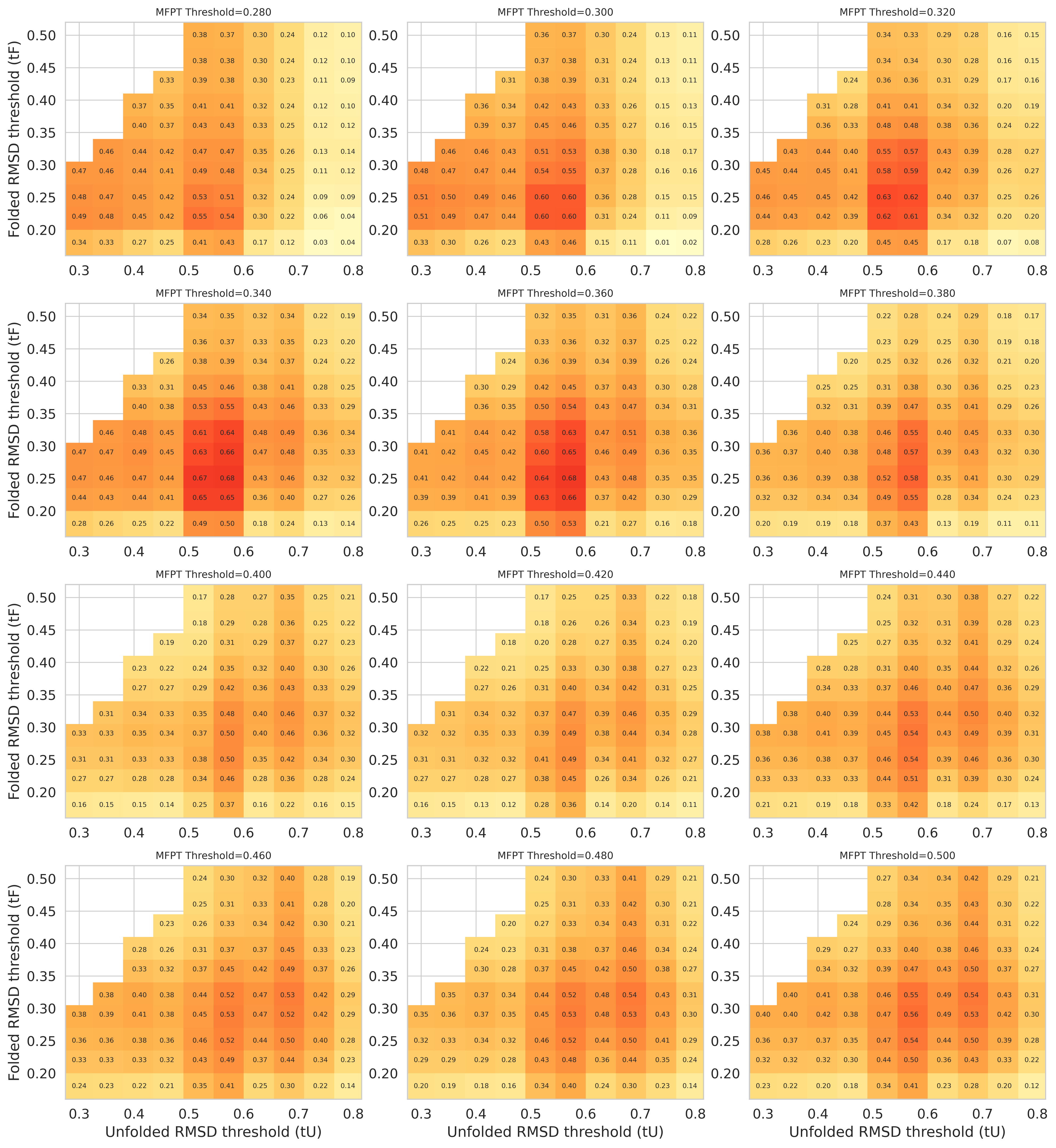}
    }
    \caption{Grid of heatmaps showing the Pearson correlation coefficient $r$ between the log MFPT ratio and the HLDA eigenvalue as a function of the RMSD thresholds used to define folded and unfolded states ($t_F$, $t_U$). Each panel corresponds to a different RMSD defining unfolding events ($t_{\mathrm{FPT}}$), with color indicating correlation strength. Correlations persist over a range of values, demonstrating robustness of the HLDA--kinetics relationship to state definition.}

    \label{fig:hlda:mfpt:heatmap}
\end{figure}

\begin{figure}[t]
    \centering
    \makebox[\textwidth][c]{%
        \includegraphics[width=1.1\textwidth]{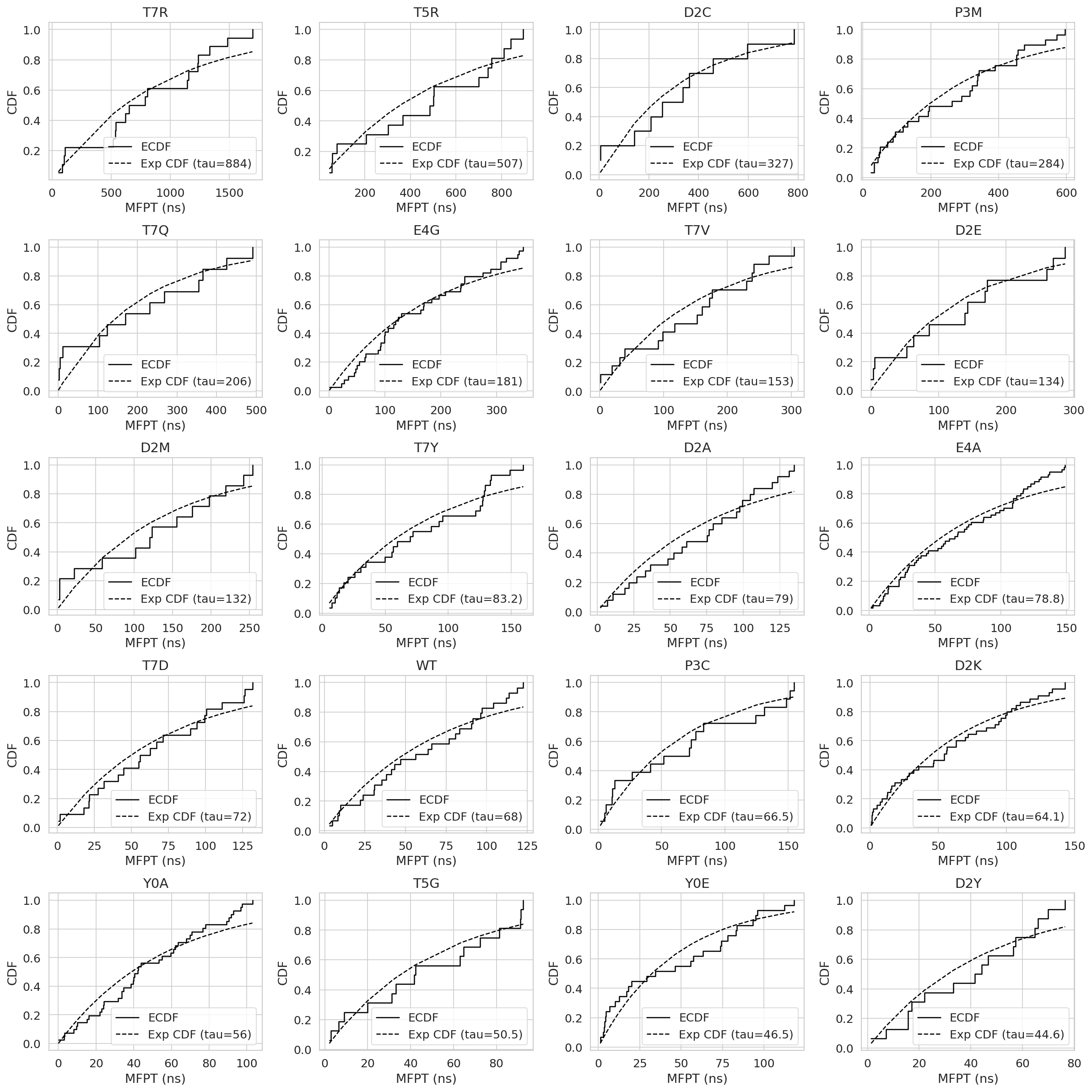}
    }
    \caption{Empirical cumulative distribution function (ECDF) of first-passage times aggregated over all mutants, sorted from slowest to fastest unfolding, compared with the theoretical exponential CDF. The agreement supports an exponential first-passage-time model for the unfolding process.}

    \label{fig:ecdf:mutants}
\end{figure}

\begin{figure}[t]\ContinuedFloat
    \centering
    \makebox[\textwidth][c]{%
        \includegraphics[width=1.1\textwidth]{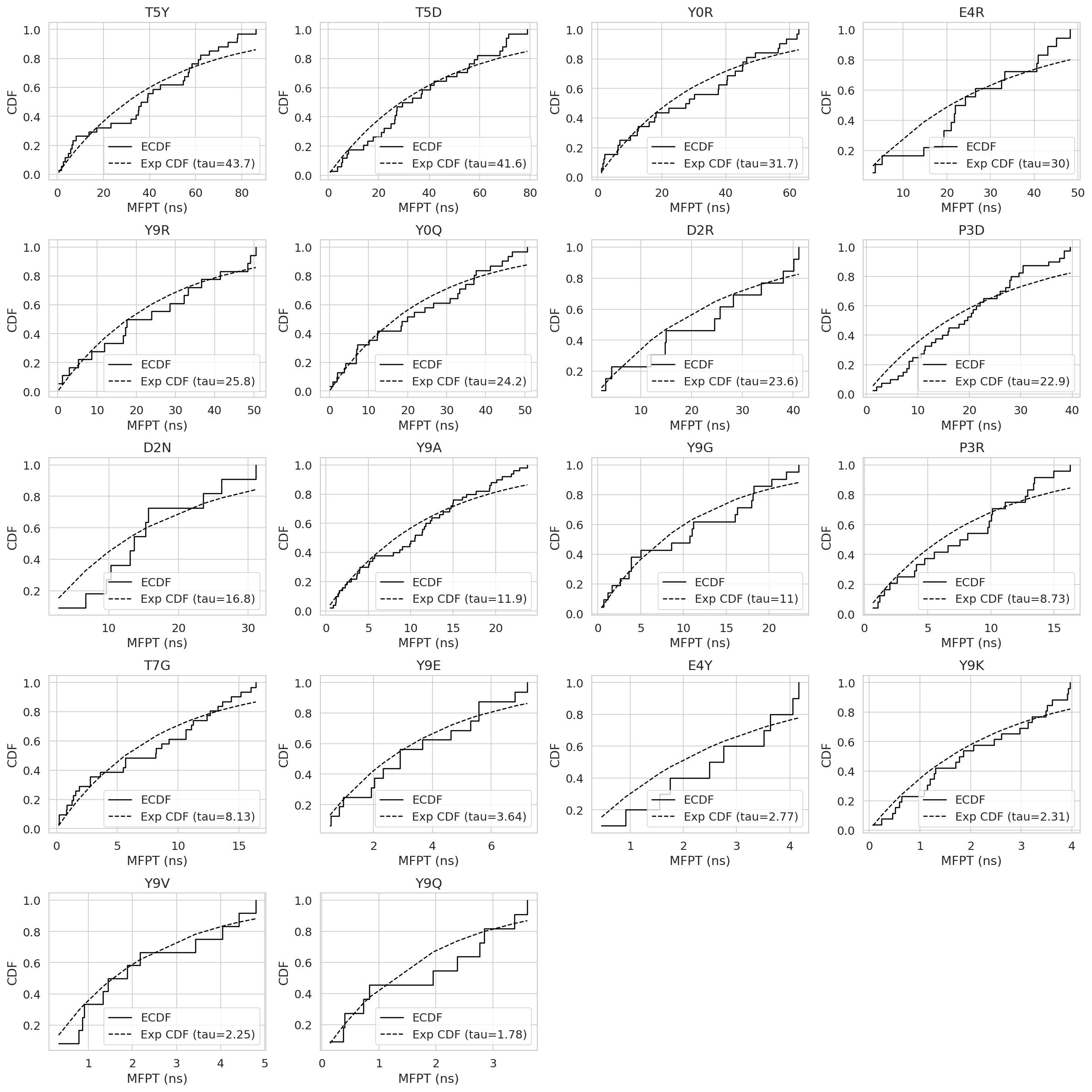}
    }
    \caption{(continued)}
    \label{fig:ecdf:mutants:cont}
\end{figure}

\end{document}